\documentclass[pra,twocolumn,amsmath,amssymb,floatfix,reprint,footinbib,superscriptaddress,longbibliography,showkeys]{revtex4-1}
\usepackage{dcolumn}
\usepackage{graphicx}
\usepackage{mathrsfs}
\usepackage{mdwlist}
\usepackage{subfigure}
\usepackage{booktabs}
\usepackage{multirow}
\usepackage{amsmath}
\usepackage{upgreek}
\usepackage{dsfont}
\usepackage{amstext}
\usepackage{amssymb}
\usepackage{amsbsy}
\usepackage{appendix}
\usepackage{soul}
\usepackage{threeparttable}
\usepackage{bbm}
\usepackage{amsthm}
\usepackage{graphicx}
\usepackage{lipsum}
\usepackage{xcolor}
\usepackage[none]{hyphenat}
\usepackage{textcomp}
\usepackage{color}
\usepackage[colorlinks,citecolor=blue]{hyperref}
\definecolor{Dgreen}{RGB}{0, 100, 0}
\usepackage{url}
\usepackage[colorlinks]{hyperref}
\hypersetup{%
	plainpages=true,
	breaklinks=true,  %not default in dvips mode, so we must specify
	hypertexnames=false,  %not ideal, but needed when pagenums duplicate (`i' vs. `1')
	pageanchor=true,
	colorlinks=true,
	linkcolor={blue},
	citecolor={red},
	urlcolor={blue},
	anchorcolor={black}
}

\begin{document}
%\begin{CJK*}{UTF8}{gbsn}

\title{Fault-tolerant multiqubit geometric entangling gates using photonic cat-state qubits}
\author{Ye-Hong Chen}
\affiliation{Theoretical Quantum Physics Laboratory, RIKEN Cluster for Pioneering Research, Wako-shi, Saitama 351-0198, Japan}
\affiliation{RIKEN Center for Quantum Computing (RQC), Wako-shi, Saitama 351-0198, Japan}

\author{Roberto Stassi}
\affiliation{Theoretical Quantum Physics Laboratory, RIKEN Cluster for Pioneering Research, Wako-shi, Saitama 351-0198, Japan}
\affiliation{Dipartimento di Scienze Matematiche e Informatiche, Scienze Fisiche e Scienze della Terra, Universit\`{a} di Messina, 98166, Messina, Italy}

\author{Wei Qin}
%\thanks{wei.qin@riken.jp}
\affiliation{Theoretical Quantum Physics Laboratory, RIKEN Cluster for Pioneering Research, Wako-shi, Saitama 351-0198, Japan}

\author{Adam Miranowicz}
\affiliation{Theoretical Quantum Physics Laboratory, RIKEN Cluster for Pioneering Research, Wako-shi, Saitama 351-0198, Japan}
\affiliation{Institute of Spintronics and Quantum Information,
	Faculty of Physics, Adam Mickiewicz University, 61-614 Pozna\'n, Poland}

\author{Franco Nori}%\thanks{fnori@riken.jp}
\affiliation{Theoretical Quantum Physics Laboratory, RIKEN Cluster for Pioneering Research, Wako-shi, Saitama 351-0198, Japan}
\affiliation{RIKEN Center for Quantum Computing (RQC), Wako-shi, Saitama 351-0198, Japan}
\affiliation{Department of Physics, University of Michigan, Ann Arbor, Michigan 48109-1040, USA}

\date{\today}

\begin{abstract}
  We propose a theoretical protocol to implement multiqubit geometric gates (i.e., the M{\o}lmer-S{\o}rensen gate) using photonic cat-state qubits. 
  These cat-state qubits stored in high-$Q$ resonators are promising for hardware-efficient 
  universal quantum computing. Specifically, in the limit of strong two-photon drivings, phase-flip errors of the cat-state qubits 
  are effectively suppressed, leaving only a bit-flip error to be corrected.
  {Because this dominant error commutes with the evolution operator, our protocol preserves the error bias, and, thus, can lower the code-capacity threshold for error correction.}
  A geometric evolution guarantees the robustness of the protocol
  against stochastic noise along the evolution path. 
  Moreover, by changing detunings of the cavity-cavity couplings at a proper time,
  the protocol can be robust against parameter imperfections (e.g., the total evolution time)
  without introducing extra noises into the system. As a result, the gate can produce multi-mode entangled cat states
  in a short time with high fidelities. 
\end{abstract}

%\pacs {03.67.}

\maketitle
\section{Introduction}
Quantum computers promise to drastically outperform classical computers on certain problems, such as factoring and unstructured database searching \cite{HidaryBook,LiptonBook,Nielsen2000,Kockum2019,Kjaergaard2020}. Recent experiments with superconducting
qubits \cite{Arute2019} and photons \cite{Zhong2020}
have already demonstrated quantum advantage.
To perform useful large-scale
quantum computation, fragile quantum states must be protected from errors, which
arise due to their inevitable interaction with the environment \cite{HidaryBook,LiptonBook,Nielsen2000}. Aiming at this problem, 
strategies for quantum error correction have been developed in the past decades \cite{Shor1995,Steane1996,Kitaev2003,Gottesman2001,Mirrahimi2014,Mirrahimi2016,Chamberland2020,Cai2021,Ma2021,Ralph2003,Gilchrist2004,Gaitan2008,Daniel2009,gottesman2010introduction,Kjaergaard2020,Fowler2012,Devitt2013,Zhang2018,PuriPrx2019,Litinski2019}.
For instance, because most noisy environments
are only locally correlated, quantum
information can be protected by employing nonlocality using, e.g., entangled qubit states \cite{Shor1995}, spatial distance \cite{Kitaev2003}, and their
combinations \cite{Fowler2012,Litinski2019}. Note that this strategy has been extended to states that are nonlocal in the phase space
of an oscillator \cite{Kjaergaard2020,Mirrahimi2016,Chamberland2020,Cai2021,Ma2021,Gottesman2001,Ralph2003,Gilchrist2004,Mirrahimi2014,Albert2016,Michael2016,Heeres2017,Li2017,Chou2018,Rosenblum2018,Albert2019,Xu2020,Gertler2021}, such as Schr\"{o}dinger cat states \cite{Dodonov1974,Ralph2003,Gilchrist2004,Liu2005,Kira2011,Gribbin2013,Leghtas2015,Chen2021,Chen2020,Stassi2021}. 
Encoding quantum information in such bosonic states has the benefit of involving fewer physical components.
{ In particular, cat-state qubits (which are formed by even and odd coherent states of a single optical mode) are promising for
hardware-efficient universal quantum computing because these cat-state qubits are noise biased \cite{Cai2021,Ma2021}.
This kind of logical qubit experiences only bit-flip noise, while the phase-flip errors are exponentially suppressed.}
Additional layers
of error correction can focus only on the
bit-flip error, so that the number of building blocks can be significantly reduced \cite{PuriPrx2019,Guillaud2019,Gottesman2001}.
%which are
%superpositions of coherent states.
%The cat-state qubits are attractive because coherent states
%are eigenstates of the photon annihilation operator and
%therefore single-photon loss induces simple, tractable errors \cite{Heeres2017,Ofek2016}.

%As a result, one can apply efficient error-protection into the physical layer while maintaining
%simplicity.

In this manuscript, we propose to use Kerr cat-state qubits to implement 
multiqubit geometric gates, i.e., the well-known M{\o}lmer-S{\o}rensen (MS) entangling gate \cite{Sorensen1999,Sorensen2000} and
its multiqubit generalizations \cite{Molmer1999}.
Generally, the MS gate is a two-qubit geometric gate 
possessing
a built-in noise-resilience feature against certain types
of local noises \cite{Solinas2004,Zhu2005,Zheng2004,Zheng2016,Song2017,Xue2017,Kang2018}.
It is also a significant resource for Grover's quantum search algorithm \cite{Nielsen2000,Grover1997} without a third ancilla bit \cite{Brickman2005}.

Previous works \cite{Haljan2005,Kirchmair2009,HayesPRL2012,Lemmer2013,Haddadfarshi2016,Takahashi2017,Shapira2018,Manovitz2017,Mitra2020,Wang2021} implementing the MS gate using physical qubits (such as trapped ions and atoms) may experience various errors including bit flips, phase flips, qubit dephasing, etc.
Thus, a huge physical resource is needed to correct the various errors \cite{Haffner2008,Bruzewicz2019,HayesPRL2012,ParradoRodrguez2021}.
This requirement has driven researchers to optimize such implementations with respect to
speed and robustness to nonideal control environments using extra
control fields \cite{Manovitz2017,Shapira2018,Mitra2020,Wang2021}. However, additional control fields
may induce extra noises that should be corrected by using additional physical resources.
All the above factors impede in scaling up the number of qubits because error channels increase 
when the number of physical qubits increases.

Instead, Kerr cat-state qubits, which 
experience only a bit-flip error, can be an excellent choice to overcome
the above problems. 
{This is because the dominant error commutes with the MS gate matrix. As a result, an erroneous gate
operation is equivalent to an error-free gate followed by an error, i.e., our cat-code gates preserve the error bias.
The code capacity threshold for error correction using such biased-noise qubits is higher than that using qubits without such structured noise \cite{Guillaud2019,Puri2020}.
We suggest, using cavity and circuit quantum electrodynamics \cite{PuriPrx2019,Puri2017,Chen2021}, to realize our protocol. 
{This can avoid some problems in trapped-ion implementations, such as the limitation of a Lamb-Dicke parameter. Note that the MS entangling gate was initially proposed \cite{Sorensen1999,Sorensen2000} and experimentally realized \cite{Haljan2005} in trapped-ion systems.}

\begin{figure}
	\centering
	\scalebox{1.05}{\includegraphics{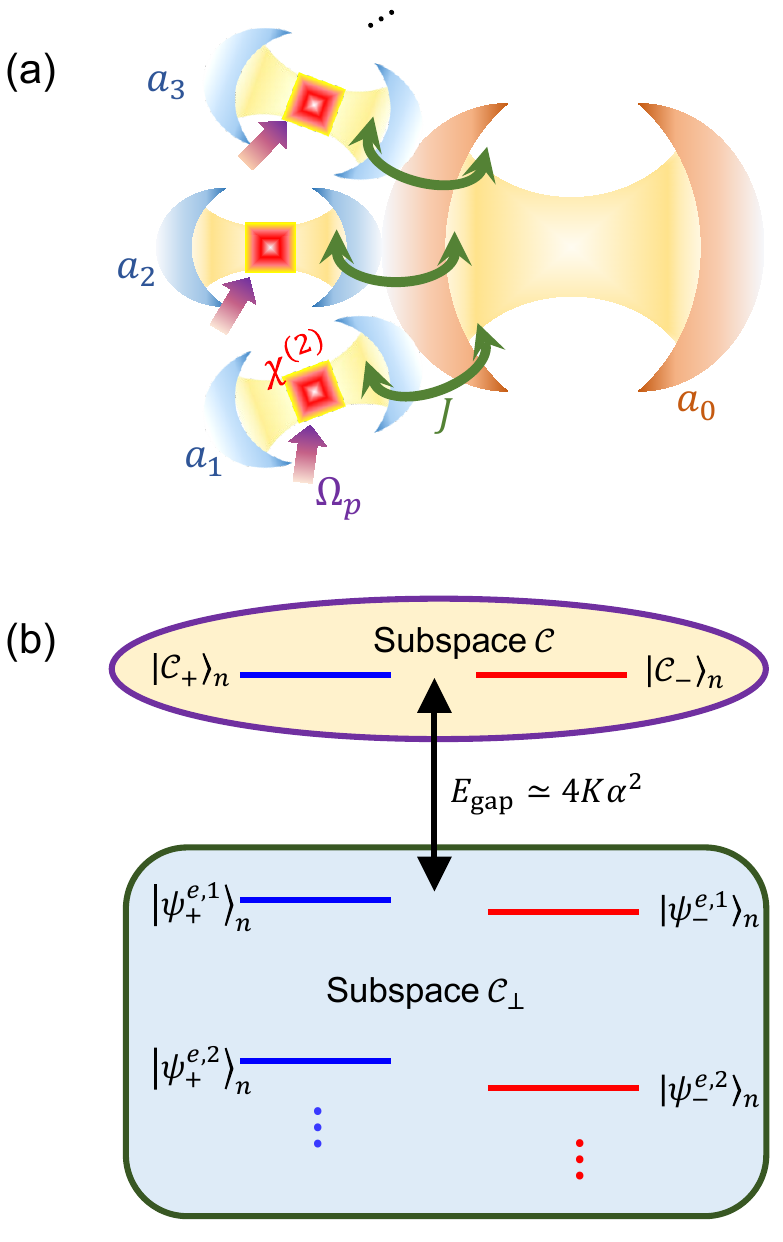}}
	\caption{{(a) Schematic of $N$ Kerr-nonlinear resonators coupled to another resonator. Driving the $\chi^{(2)}$ nonlinearity induces
		a two-photon driving in the cavity mode $a_{n}$. 
		(b) Eigenspectrum of the $n$th Kerr parametric
		oscillator, $H_{n}^{\rm{Kerr}}$, in the rotating frame determined by Eq.~(\ref{R16}).
		The excited states appear at a lower energy because the Kerr nonlinearity is negative.
	}}
	\label{fig1}
\end{figure}

\section{Model and effective Hamiltonian}
We consider $N$ Kerr-nonlinear resonators ($a_{1}$, $a_{2}$, $\ldots$, $a_{N}$) with the same frequency $\omega_{c}$, which are simultaneously coupled to 
another resonator ($a_{0}$) with frequency $\omega_{0}$ [see Fig.~\ref{fig1}(a)]. The interaction Hamiltonian is  
\begin{align}
	H_{\rm{int}}=\sum_{n=1}^{N}Ja_{n}a_{0}^{\dag}\exp{(i\Delta t)}+{\rm{h.c.}},
\end{align} 
where $J$ is the intercavity coupling strength and $\Delta=\omega_{0}-\omega_{c}$ is the detuning. Hereafter, we assume that $\hbar=1$. 
Each Kerr-nonlinear resonator is resonantly driven by a two-photon drive of frequency $\omega_{p}=2\omega_{c}$
and amplitude $\Omega_{p}$ \cite{PuriPrx2019,Puri2017}.
The total Hamiltonian of the system in the interaction picture reads
\begin{align}\label{eq1}
   H=&\sum_{n=1}^{N}H_{n}^{\rm{Kerr}}+H_{\rm{int}},	\cr
   H_{n}^{\rm{Kerr}}=&-Ka_{n}^{\dag 2}a_{n}^{2}+\left(\Omega_{p}a_{n}^{2}+{\rm{h.c.}}\right),
\end{align} 
where $H_{n}^{\rm{Kerr}}$ describes Kerr parametric
oscillators (KPOs) with Kerr nonlinearity $K$ \cite{Miranowicz2016,Blais2012}.

To understand the Hamiltonian $H_{n}^{\rm{Kerr}}$ in Eq.~(\ref{eq1}), {following Refs.~\cite{Puri2017,PuriPrx2019,Puri2020,Cai2021,Ma2021}},
we can apply the displacement transformation 
\begin{align}
	D_{n}(\pm\alpha)=\exp\left[\pm\alpha \left(a_{n}^{\dag}-a_{n}\right)\right],
\end{align} 
so that Eq.~(\ref{eq1}) becomes
\begin{align}\label{R16}
	H'_{n}=&D_{n}(\pm\alpha)H_{n}^{\rm{Kerr}}D_{n}^{\dag}(\pm\alpha)\cr
	       =&-K\left[4\alpha^{2}a_{n}^{\dag}a_{n}-a_{n}^{\dag 2}a_{n}^{2}\mp 2\alpha(a_{n}^{\dag 2}a_{n}+{\rm{h.c.}})\right].
\end{align}
Hereafter, we choose $\left\{K,\Omega_{p},J,\Delta\right\}>0$ for simplicity, then, $\alpha=\alpha^{*}>0$.
Because of $H'_{n}|\nu=0\rangle=0$, the vacuum state $|0\rangle$ is exactly an eigenstate of $H'_{n}$.
Therefore, the coherent states $|\pm\alpha\rangle$, or, equivalently, their superpositions
\begin{align}
	 |\mathcal{C}_{\pm}\rangle_{n}=\mathcal{N}_{\pm}\left[D_{n}(\alpha)\pm D_{n}(-\alpha)\right]|0\rangle_{n},
\end{align}
are the eigenstates of $H_{n}^{\rm{Kerr}}$ in the original frame.
Here, $\mathcal{N}_{\pm}$ are normalization coefficients.
In the limit of large $\alpha$, $\alpha^{2}\gg\alpha^{1},\alpha^{0}$, 
Eq. (\ref{R16}) is approximated by 
\begin{align}
	H'_{n}\simeq -4K\alpha^{2}a_{n}^{\dag}a_{n},
\end{align} which is
the Hamiltonian of a (inverted) harmonic oscillator \cite{PuriPrx2019}.
Thus, in the original frame, the eigenspectrum of $H_{n}^{\rm{Kerr}}$ can be divided into even- and odd-parity
manifolds as shown in Fig.~\ref{fig1}(b).
The excited states appear at a lower energy because the Kerr nonlinearity is negative.
{In the limit of large $\alpha$,  we can approximatively express the first-excited
	states as the two orthogonal states}
\begin{align}
	|\psi_{\pm}^{e,1}\rangle_{n}=\mathcal{N}_{e}^{\pm}\left[D_{n}(\alpha)\mp D_{n}(-\alpha)\right]|\nu=1\rangle_{n},
\end{align} which are the even- and odd-parity states,
respectively. Here, $\mathcal{N}_{e}^{\pm}$ are normalization coefficients.

{As shown in Fig.~\ref{fig1}(b), the orthogonal cat states $|\mathcal{C}_{\pm}\rangle_{n}$ can 
span a cat subspace $\mathcal{C}$,} which is separated from the excited eigenstates of KPO by an energy gap $E_{\rm{gap}}\simeq4K\alpha^{2}$ (i.e., the energy gap between $|\mathcal{C}_{\pm}\rangle_{n}$ and $|\psi_{\pm}^{e,1}\rangle_{n}$).
In the limit of large $\alpha$,
the action of $a_{n}$ only flips the two cat states, i.e., 
\begin{align}
	 a_{n}|\mathcal{C}_{\pm}\rangle_{n}\simeq\alpha|\mathcal{C}_{\mp}\rangle_{n}.
\end{align}
The action of $a_{n}^{\dag}$ on a state in
the cat subspace causes transitions 
to the excited states, i.e., 
\begin{align}
	a_{n}^{\dag}|\mathcal{C}_{\pm}\rangle_{n}\rightarrow \alpha|\mathcal{C}_{\mp}\rangle_{n}+|\psi_{\mp}^{e,1}\rangle_{n}.
\end{align}
{When the KPOs are coupled to the cavity mode $a_{0}$, with the interaction Hamiltonian $H_{\rm{int}}$, the Hamiltonian describing transitions to the excited states (projected onto the eigenstates of $H_{n}^{\rm{Kerr}}$) is
\begin{align}\label{eqR5}
	H_{e}=&\sum_{n=1}^{N} \frac{E_{\rm{gap}}}{2}(|\mathcal{C}_{\pm}\rangle_{n}\langle \mathcal{C}_{\pm}|-|\psi_{\pm}^{e,1}\rangle_{n}\langle\psi_{\pm}^{e,1}|)\cr
	                &+J\left[|\psi_{\mp}^{e,1}\rangle_{n}\langle \mathcal{C}_{\pm}|a_{0}\exp{(-i\Delta t)}+{\rm{h.c.}}\right].
\end{align}
Here, we have defined and used the projection operator
\begin{align}
	P_{\rm{KPO}}=\sum_{n=1}^{N}\left(|\mathcal{C}_{\pm}\rangle_{n}\langle\mathcal{C}_{\pm}|+\sum_{\nu=1}^{\infty}|\psi_{\pm}^{e,\nu}\rangle_{n}\langle \psi_{\pm}^{e,\nu}|\right).
\end{align}
Because $E_{\rm{gap}}>0$, according to Eq.~(\ref{eqR5}), the probability of excitation
	to the states $|\psi_{\pm}^{e,1}\rangle_{n}$ is suppressed by 	
\begin{align}
 P_{e}\sim\frac{NJ^2}{(E_{\rm{gap}}+\Delta)^2},
\end{align}
{which is proportional to both, the number $N$ of cat-state qubits and the square of the coupling strength $J$.}
Therefore, in the limit of $J\ll E_{\rm{gap}}$, the excited eigenstates of the KPOs remain unpopulated. Then, the dynamics of the system is restricted
in the cat subspace with an effective Hamiltonian
\begin{eqnarray*}
	H_{\rm{eff}}&\simeq&\sum_{n=1}^{N}\frac{\Omega_{p}^{2}}{K}\left(|\mathcal{C}_{-}\rangle_{n}\langle \mathcal{C}_{-}|+|\mathcal{C}_{+}\rangle_{n}\langle \mathcal{C}_{+}|\right)\cr\cr
	&&+J\alpha\left[|\mathcal{C}_{+}\rangle_{n}\langle\mathcal{C}_{-}|\left(a_{0}e^{-i\Delta t}+a_{0}^{\dag}e^{i\Delta t}\right)+{\rm{h.c.}}\right].
\end{eqnarray*}
Here, the first-line expression in $H_{\rm{eff}}$ can be dropped because it is proportional to the identity matrix $\mathbbm{1}_{n}=|\mathcal{C}_{-}\rangle_{n}\langle \mathcal{C}_{-}|+|\mathcal{C}_{+}\rangle_{n}\langle \mathcal{C}_{+}|$
in the dressed-state subspace.
In the limit of large $\alpha$, by using the definition of Pauli matrices $\sigma_{n}^{+}=|\mathcal{C}_{-}\rangle_{n}\langle\mathcal{C}_{+}|$ and $\sigma_{n}^{-}=\left(\sigma_{n}^{+}\right)^{\dag}$, 
$H_{\rm{eff}}$ becomes
\begin{align}
	H_{\rm{eff}}\simeq& J\alpha\sum_{n=1}^{N}\left[\left(\sigma_{n}^{+}+\sigma_{n}^{-}\right)\left(a_{0}e^{-i\Delta t}+a_{0}^{\dag}e^{i\Delta t}\right)\right]\cr
	=&2J\alpha S_{x}\left(a_{0}e^{-i\Delta t}+a_{0}^{\dag}e^{i\Delta t}\right),
\end{align}
where 
$S_{x}=\frac{1}{2}\sum_{n}(\sigma_{n}^{+}+\sigma_{n}^{-})$.
}

\section{Implementing the MS gates}
The integral of $H_{\rm{eff}}$ can be calculated exactly \cite{Sorensen2000},
\begin{eqnarray*}\label{eq3}
  U_{\rm{MS}}(t)=%e^{-iS_{x}[F(t)x_{0}+G(t){0}]}e^{-iA(t)S_{x}^{2}},
  \exp{\left\{-i\left[\chi(t)a_{0}^{\dag}S_{x}+{\rm{h.c.}}\right]\right\}}\exp{\left[-i\beta(t)S_{x}^{2}\right]},
\end{eqnarray*} 
where 
\begin{align}
\chi(t)=&\frac{2iJ\alpha}{\Delta}\left[1-\exp{(i\Delta t)}\right], \cr
\beta(t)=&\left(\frac{2J\alpha}{\Delta}\right)^{2}\left(\sin\Delta t-\Delta t\right).
\end{align}
One observes that in the phase space determined by the cavity mode $a_{0}$, $\chi(t)$ draws $m$ circles with a radius $r=2J\alpha/\Delta$ and a rotation angle $\theta=\Delta t_{g}$ when 
\begin{align}
	t=t_{g}=\frac{2m\pi}{\Delta}.\ \ \ \ (m=1,2,\ldots)
\end{align}
Here, $t_{g}$ is the gate time.
Thus,
the cavity mode $a_{0}$ evolves along a
circle in phase space and returns to its (arbitrary) initial state after $m$ periods.
Meanwhile, $\beta(t_{g})$ can be expressed by the area $A$ enclosed by $\chi(t)$ as
\begin{align}
	\beta(t_{g})=-2m\pi r^2=-2mA,
\end{align} 
which is a geometric phase.
The evolution operator at the time $t=t_{g}$ reads
\begin{align}\label{eq4}
  U_{\rm{MS}}(t_{g})=\exp\left[-i\beta(t_{g}) S_{x}^{2}\right].
\end{align}
%When choosing $\Delta=4\sqrt{m}J\alpha$, the accumulated Berry phase is $\beta(\tau_{m})=-\pi/2$.

In particular, when $\beta(t_{g})=-\pi/2$ and $N$ is even,
$U_{\rm{MS}}(t_{g})$ accomplishes the 
transformations, i.e.,
\begin{eqnarray*}
	&&\bigotimes_{n=1}^{N}|\mathcal{C}_{\pm}\rangle_{n}~\xrightarrow{U_{\rm{MS}}(t_{g})}~\frac{1}{\sqrt{2}}\left(\bigotimes_{n=1}^{N}|\mathcal{C}_{\pm}\rangle_{n}+i\bigotimes_{n=1}^{N}|\mathcal{C}_{\mp}\rangle_{n}\right),
	%|\mathcal{C}_{-}\rangle_{1}|\mathcal{C}_{\pm}\rangle_{2}\rightarrow \frac{1}{\sqrt{2}}\left(|\mathcal{C}_{-}\rangle_{1}|\mathcal{C}_{\pm}\rangle_{2}+i|\mathcal{C}_{\mp}\rangle_{1}|\mathcal{C}_{+}\rangle_{2}\right),
\end{eqnarray*}
which maps product states (i.e., the input state $|\psi_{\rm{in}}\rangle$) into maximally
entangled cat states (i.e., the output state $|\psi_{\rm{out}}\rangle$). 
Accompanied
by single-qubit rotations \cite{Grimm2020,Masluk2012,Pop2014,Cohen2017}, the MS gate can be 
applied in Grover's quantum search algorithm for
both the marking and state amplification steps \cite{Brickman2005,DiCarlo2009}.
A possible approach for such single-qubit gates is shown in Appendix~\ref{App2}.
The generation of input states in a KPO has been experimentally realized \cite{Grimm2020}.
For instance, using time-dependent two-photon drivings, a cat state with fidelity $\gtrsim 95\%$ \cite{Puri2017} in the presence of decoherence can be generated.
For clarity, in Appendix~\ref{App1}, we describe a possible protocol to generate the cat states.
{Hereafter, we use QuTip \cite{Qutip1,Qutip2} for numerical simulations.}

\begin{figure}
	\centering
	\scalebox{0.92}{\includegraphics{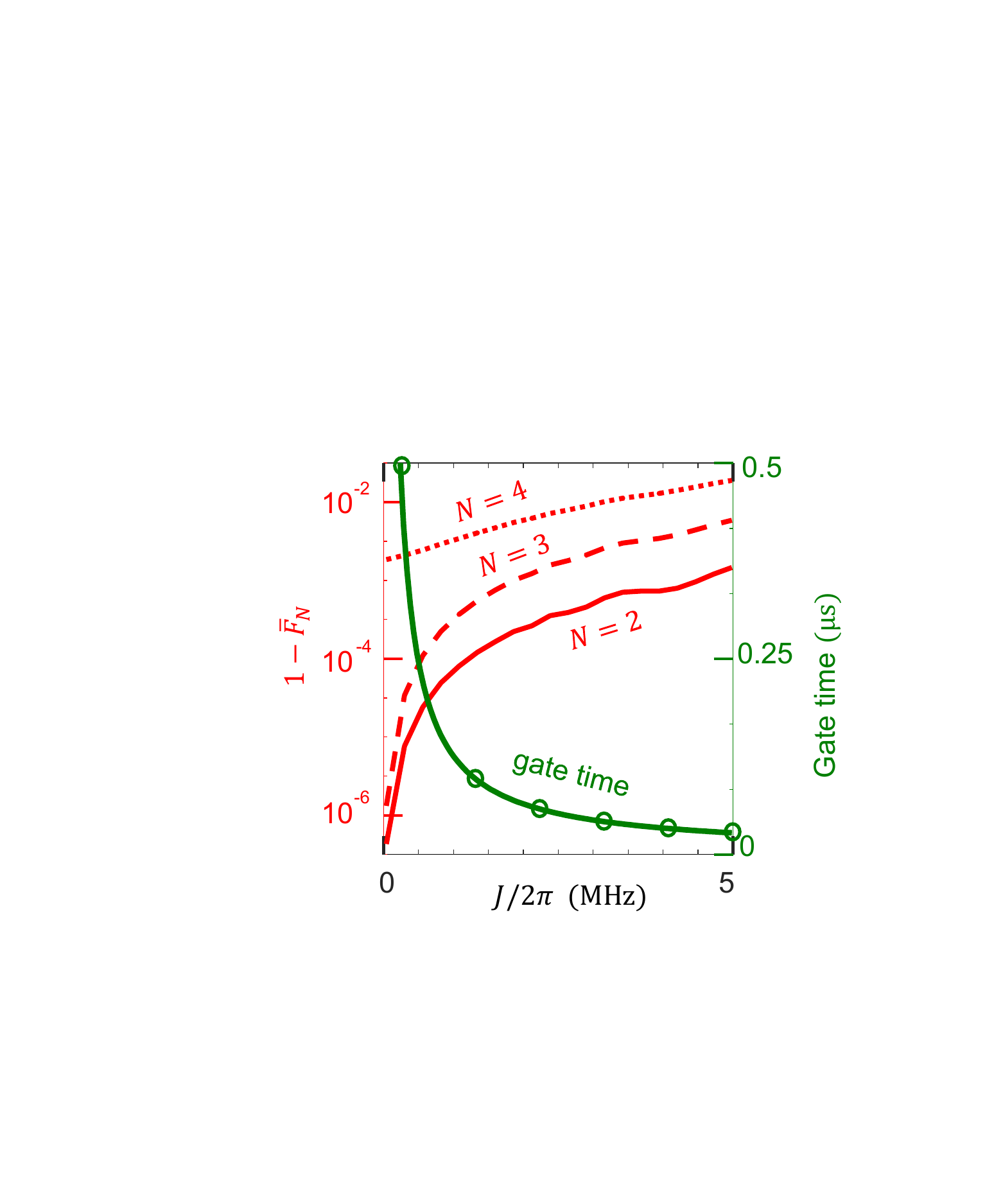}}
	\caption{Gate infidelities ($1-\bar{F}_{N}$) and gate time $t_{g}=\pi/(2J\alpha)$ (i.e., $m=1$) calculated for the total Hamiltonian $H$,
			for different values of the coupling strength $J$. Here, the gate fidelity $\bar{F}_{N}$ is defined in Eq.~(\ref{eq17}). We assume a realistic Kerr nonlinearity $K/2\pi=5~{\rm{MHz}}$ \cite{WangPrx2019}.
			For $\beta(t_{g})=-\pi/2$, we choose other parameters
			$\Delta=4J\alpha$ [i.e., $m=1$ in Eq.~(\ref{eq20})] and $\alpha=2$. 
	}
	\label{fig1a}
\end{figure}

The average fidelity of an $N$-qubit gate
over all possible initial states is defined by \cite{Zanardi2004,Pedersen2007}
\begin{align}\label{eq17}
\bar{F}_{N}=&\frac{\mathrm{Tr}(MM^\dag)
+|\mathrm{Tr}(M)|^2}{D^2+D}, \cr
M=&\mathcal{P}_{{c}}U^\dag_{\rm{MS}}U(t_{g})\mathcal{P}_{{c}}.
\end{align}
Here, $\mathcal{P}_{{c}}$ ($D$) is the projector
(dimension) of the computing subspace,
 {and 
 \begin{align}
 	U(t_{g})=\exp{\left(-iH t_{g}\right)}
 \end{align} is the actual evolution operator of the system calculated from the total Hamiltonian $H$.
Unless specified otherwise, the numerical simulations in our manuscript are carried out using the full Schr\"{o}dinger equation (for coherent dynamics) and the full Lindblad master equation (for incoherent dynamics) with the full Hamiltonian {$H$ in Eq.~(\ref{eq1})} in the entire space.
%Hereafter, we refer $\bar{F}_{N}$ as ``gate fidelity''.

Current
experiments using superconducting systems \cite{WangPrx2019,Grimm2020,Leghtas2015,Touzard2019,Gao2018} have achieved a driving amplitude $\Omega_{p}/2\pi\sim 10$--$40~{\rm{MHz}}$ and 
a Kerr nonlinearity $K/2\pi\sim 1$--$10~\rm{MHz}$.
Hereafter, we choose $K/2\pi=5~{\rm{MHz}}$.
Note that
$\Delta$ and $J$ should obey 
\begin{align}\label{eq20}
	\Delta=4\sqrt{m}J\alpha,
\end{align}
for $\beta(t_{g})=-{\pi}/2$.
Therefore, the gate time 
\begin{align}
	t_{g}=\frac{\pi\sqrt{m}}{2J\alpha}
\end{align} 
is inversely proportional to 
$J$ [see the green-solid curve with circles in Fig.~\ref{fig1a}].
The gate infidelities ($1-\bar{F}_{N}$) for $N=2,~3,~4$ versus $J$ are shown in  
Fig.~\ref{fig1a}. When $J/2\pi\lesssim 0.5~{\rm{MHz}}$,
we can achieve high-fidelity $\bar{F}_{N}\gtrsim99.9\%$ multiqubit gates within a gate time $t_{g} \lesssim500~{\rm{ns}}$. 
%Increasing the detuning $\Delta$ can further increase the gate fidelity to $\gtrsim99.99\%$, but it leads to a longer gate time, which increases the infidelities in the presence of decoherence.
 }

\section{Analysis of decoherence}
{For the resonators, we consider two kinds of noise: single-photon loss and pure dephasing.
The system dynamics is described by
the Lindblad master equation 
\begin{align}\label{eq4a}
  \dot{\rho}=-i[H,\rho]+\sum_{j=0}^{N}\kappa_{j} \mathcal{D}[a_{j}]\rho+\gamma_{j}\mathcal{D}[a_{j}^{\dag}a_{j}]\rho,
\end{align}
where $\mathcal{D}[o]\rho=o\rho o^{\dag}-\left(o^{\dag}o\rho+\rho o^{\dag}o\right)/2$
is the standard Lindblad superoperator and $\kappa_{j}$ ($\gamma_{j}$) is the single-photon loss (pure dephasing) rate of the $j$th cavity mode. Without loss of generality, for the KPOs, we assume $\kappa_{n}=\kappa$ and $\gamma_{n}=\gamma$ ($n=1,2,\ldots,N$). 
Note that the influence of decoherence in the cavity mode $a_{0}$ is different
from that in the KPOs. We initially consider only decoherence in the cavity mode $a_{0}$ i.e., assume that $\kappa_{n}=\gamma_{n}=0$.}
For simplicity, we choose an initial state 
\begin{align}
  |\psi_{\rm{in}}\rangle=|0\rangle_{0}\bigotimes_{n=1}^{N}|\mathcal{C}_{+}\rangle_{n}.
\end{align}
The fidelity 
\begin{align}\label{eq23}
F_{\rm{out}}=\langle\psi_{\rm{out}}|\rho(t_{g})|\psi_{\rm{out}}\rangle,
\end{align} 
of the output state
versus decoherence in the cavity mode $a_{0}$ is shown in Fig.~\ref{fig2}(a).
We find that the system is mostly insensitive to the decoherence of the cavity mode $a_{0}$
because it can be adiabatically eliminated for large $\Delta$.

\begin{figure*}
	\centering
	\scalebox{0.52}{\includegraphics{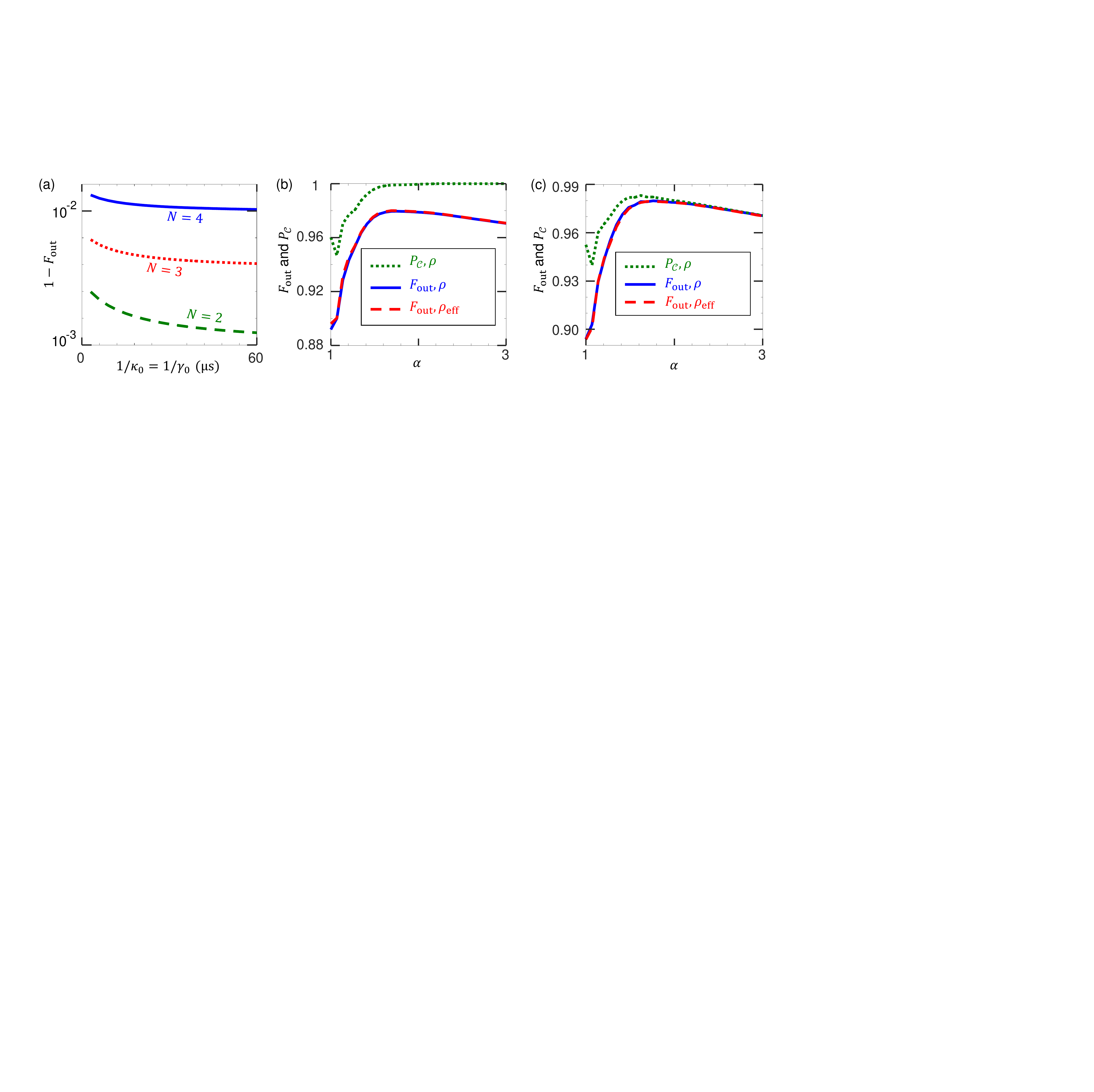}}
	\caption{{Numerical results based on the master equation in Eq.~({\ref{eq4a}}) and the effective master equation in Eq.~(\ref{eq32}). (a) Output-state infidelities $(1-{F}_{\rm{out}})$ defined in Eq.~({\ref{eq23}}) of the $N$-qubit gates ($N=2,3,4$) 
		in the presence of decoherence in the cavity mode $a_{0}$ when $\alpha=2$. }
		(b, c) Output-state fidelities $F_{\rm{out}}$ and no-leakage probability $P_{\mathcal{C}}$ of the two-qubit gate versus $\alpha$ when considering only: (b) single-photon loss $\kappa=0.1~{\rm{MHz}}$ in the KPOs and (c) pure dephasing $\gamma=0.1~{\rm{MHz}}$
		in the KPOs. {Note that the blue-solid and green-dotted curves in (b) and (c) are plotted using the Lindblad master equation in Eq.~(\ref{eq4a}).
		The red-dashed curves in (b) and (c) are plotted for the effective Lindblad master equation in Eq.~(\ref{eq32}).}
		Other parameters are
		$J/2\pi=K/2\pi=5~{\rm{MHz}}$ and $\Delta=4J\alpha$ [i.e., $m=1$ in Eq.~(\ref{eq20})], resulting in a gate time $t_{g}=25~{\rm{ns}}$. 
		For clarity, when studying one kind of error, we assume that the other errors are zero.
	}
	\label{fig2}
\end{figure*}

For a clear understanding of the influence of decoherence in the KPOs,
we can project the system onto the eigenstates of $H_{n}^{\rm{Kerr}}$ \cite{Puri2017,PuriPrx2019,Puri2020,Cai2021,Ma2021,Scully1997Book,Agarwal2012Book}.
Then the master equation becomes
\begin{align}\label{S22}
	\dot{\rho}\simeq& -i[P_{\rm{KPO}}HP_{\rm{KPO}},\rho]+\kappa_{0}\mathcal{D}[a_{0}]\rho+\gamma_{0}\mathcal{D}[a_{0}^{\dag}a_{0}]\rho\cr
	&+\sum_{n=1}^{N}\kappa_{n} \mathcal{D}[P_{\rm{KPO}} a_{n} P_{\rm{KPO}}]\rho\cr
	&+\gamma_{n} \mathcal{D}[P_{\rm{KPO}} a_{n}^{\dag}a_{n} P_{\rm{KPO}}]\rho.
\end{align}

\begin{widetext}
\subsection{Single-photon loss in Kerr parametric oscillators}
{When $\gamma_{n}$ and $\kappa_{n}$ are much smaller than 
the energy gap $E_{\rm{gap}}$,
the dynamics
of the cat-state qubits is still well confined to the 
subspace $\mathcal{C}$ \cite{PuriPrx2019}. 
This is because a stochastic jump, corresponding to the action of $a_{n}$ on a state
in the cat-state subspace, does not cause leakage to the excited eigenstates for large $\alpha$ \cite{PuriPrx2019,Puri2020}.
We demonstrate the above conclusion in Fig.~\ref{fig2}(b), which shows the no-leakage probability 
\begin{align} 
  P_{\mathcal{C}}=\sum_{n}{}_{n}\langle \mathcal{C}_{+}|\rho(t_{g})|\mathcal{C}_{+}\rangle_{n}+{}_{n}\langle \mathcal{C}_{-}|\rho(t_{g})|\mathcal{C}_{-}\rangle_{n}\simeq 1,
\end{align}
for large $\alpha$. 
The influence of the single-photon loss in the KPOs is described by the penultimate term in Eq.~(\ref{S22}):
\begin{align}\label{R22}
	\sum_{n=1}^{N}\kappa_{n} \mathcal{D}[P_{\rm{KPO}} a_{n} P_{\rm{KPO}}]\rho 
	\approx&
	\sum_{n=1}^{N}\kappa_{n}\alpha^{2}\mathcal{D}\left[\sqrt{\tanh{\alpha^2}}|\mathcal{C}_{+}\rangle_{n}\langle\mathcal{C}_{-}|+\sqrt{\coth{\alpha^2}}|\mathcal{C}_{-}\rangle_{n}\langle\mathcal{C}_{+}|\right]\rho\cr
	&{{+\sum_{n=1}^{N}\kappa_{n}\mathcal{D}\left[\sqrt{\frac{\mathcal{N}_{+}}{\mathcal{N}_{+}^{e}}}|\mathcal{C}_{+}\rangle_{n}\langle\psi_{+}^{e,1}|+\sqrt{\frac{\mathcal{N}_{-}}{\mathcal{N}_{-}^{e}}}|\mathcal{C}_{-}\rangle_{n}\langle\psi_{-}^{e,1}|\right]\rho}}\cr
	&{+\sum_{n=1}^{N}\kappa_{n}\alpha^{2}\mathcal{D}\left[\sqrt{\frac{\mathcal{N}_{-}^{e}}{\mathcal{N}_{+}^{e}}}|\psi_{-}^{e,1}\rangle_{n}\langle\psi_{+}^{e,1}|+\sqrt{\frac{\mathcal{N}_{+}^{e}}{\mathcal{N}_{-}^{e}}}|\psi_{+}^{e,1}\rangle_{n}\langle\psi_{-}^{e,1}|\right]\rho}.
\end{align}
Here, we have omitted highly excited eigenstates of the KPOs because they are never excited in the presence of the single-photon loss.
According to the terms in the second line in Eq.~(\ref{R22}), the single-photon loss can only transfer the excited eigenstates
$|\psi_{\pm}^{e,1}\rangle_{n}$ to the ground eigenstates $|\mathcal{C}_{\pm}\rangle_{n}$. 
If a KPO is initially in cat subspace $\mathcal{C}$,
it always remains in this cat subspace in the presence of single-photon loss.
Therefore, we can neglect the terms in the last two lines in Eq.~(\ref{R22}) and obtain (for large $\alpha$)
\begin{align}\label{eq5}
	\mathcal{D}[a_{n}]\rho~\simeq~\frac{\alpha^{2}}{\sqrt{1-e^{-4\alpha^2}}} \mathcal{D}[\sigma_{n}^{x}+ie^{-2\alpha^2}\sigma_{n}^{y}]\rho,
\end{align}
where $\sigma_{n}^{x}=\sigma_{n}^{+}+\sigma_{n}^{-}$ and $\sigma_{n}^{y}=i(\sigma_{n}^{-}-\sigma_{n}^{+})$.}

{The effective master equation in Eq.~(\ref{S22}) becomes
\begin{align}\label{eq28}
		\dot{\rho}_{\rm{eff}}\simeq& -i[P_{\rm{KPO}}H P_{\rm{KPO}},\rho_{\rm{eff}}]+\kappa_{0}\mathcal{D}[a_{0}]\rho_{\rm{eff}}+\gamma_{0}\mathcal{D}[a_{0}^{\dag}a_{0}]\rho_{\rm{eff}}\cr
		&+\frac{\alpha^{2}}{\sqrt{1-e^{-4\alpha^2}}} \mathcal{D}[\sigma_{n}^{x}+ie^{-2\alpha^2}\sigma_{n}^{y}]\rho_{\rm{eff}}+\sum_{n=1}^{N}\gamma_{n} \mathcal{D}[P_{\rm{KPO}} a_{n}^{\dag}a_{n} P_{\rm{KPO}}]\rho_{\rm{eff}}.
\end{align}
}This means that in the computing subspace the single-photon loss leads primarily to a bit-flip error ($\sigma_{n}^{x}$), which is accompanied by
an exponentially small phase-flip error ($\sigma_{n}^{y}$). 
As shown in Fig.~\ref{fig2}(b), the full dynamics calculated by Eq.~(\ref{eq4a}) (blue-dotted curve) 
is in excellent agreement with the effective one using Eq.~(\ref{eq5}) (red-solid curve) for $\alpha>\sqrt{2}$. 

\subsection{Pure dephasing in Kerr parametric oscillators}
The influence of pure dephasing is described by the last term in Eq.~(\ref{S22}): 
\begin{align}\label{R24}
	\sum_{n=1}^{N}\gamma_{n}\mathcal{D}\left[P_{\rm{KPO}}a_{n}^{\dag}a_{n}P_{\rm{KPO}}\right]\rho=&\sum_{n}^{N}\gamma_{n}\alpha^4\mathcal{D}\left[\frac{\mathcal{N}_{-}}{\mathcal{N}_{+}}|\mathcal{C}_{+}\rangle_{n}\langle\mathcal{C}_{+}|+\frac{\mathcal{N}_{+}}{\mathcal{N}_{-}}|\mathcal{C}_{-}\rangle_{n}\langle\mathcal{C}_{-}|\right]\rho\cr 
	&{+\sum_{n=1}^{N}\gamma_{n}\alpha^{2}\mathcal{D}\left[\frac{\mathcal{N}_{+}}{\sqrt{\mathcal{N}_{-}\mathcal{N}_{+}^{e}}}|\psi_{+}^{e,1}\rangle_{n}\langle\mathcal{C}_{-}|+\frac{\mathcal{N}_{-}}{\sqrt{\mathcal{N}_{+}\mathcal{N}_{-}^{e}}}|\psi_{-}^{e,1}\rangle_{n}\langle\mathcal{C}_{+}|\right]\rho}\cr
	&{+\sum_{n=1}^{N}\gamma_{n}\alpha^4\mathcal{D}\left[\frac{\mathcal{N}_{-}^{e}}{\mathcal{N}_{+}^{e}}|\psi_{+}^{e,1}\rangle_{n}\langle\psi_{+}^{e,1}|+\frac{\mathcal{N}_{+}^{e}}{\mathcal{N}_{-}^{e}}|\psi_{-}^{e,1}\rangle_{n}\langle\psi_{-}^{e,1}|\right]\rho}.
\end{align}
As in the above analysis, we have ignored the highly excited eigenstates of the KPOs because they are mostly unexcited
in the evolution. 
According to the terms in the second line of Eq.~(\ref{R24}), pure dephasing 
can cause transitions from the cat states to the first-excited states with a rate $\gamma_{n}\alpha^2$.
This causes infidelities to the system. 
For large $\alpha$, we have $\mathcal{N}_{\pm}\simeq\mathcal{N}_{\pm}^{e}\simeq2$, and Eq.~(\ref{R24}) becomes (choosing $\gamma_{n}=\gamma$)
\begin{align}\label{eq6}
	\sum_{n=1}^{N}\gamma_{n}\mathcal{D}[a_{n}^{\dag}a_{n}]\rho\simeq &\gamma\sum_{n=1}^{N}\alpha^4 \mathcal{D}[\mathbbm{1}_{n}]\rho
	{+\alpha^{2}\mathcal{D}\left[\sum_{k=\pm}|\psi_{k}^{e}\rangle_{n}\langle\mathcal{C}_{k}|+{\rm{h.c.}}\right]\rho}.
\end{align}  
That is, in the computational subspace for large $\alpha$, pure dephasing
cannot cause significant infidelities.
{We can further simplify the master equation in Eq.~(\ref{eq28}) to 
\begin{align}\label{eq32}
	\dot{\rho}_{\rm{eff}}\simeq& -i[P_{\rm{KPO}}H P_{\rm{KPO}},\rho_{\rm{eff}}]+\kappa_{0}\mathcal{D}[a_{0}]\rho_{\rm{eff}}+\gamma_{0}\mathcal{D}[a_{0}^{\dag}a_{0}]\rho_{\rm{eff}}\cr
	&+\frac{\alpha^{2}}{\sqrt{1-e^{-4\alpha^2}}} \mathcal{D}[\sigma_{n}^{x}+ie^{-2\alpha^2}\sigma_{n}^{y}]\rho_{\rm{eff}}+\gamma\sum_{n=1}^{N}\alpha^4 \mathcal{D}[\mathbbm{1}_{n}]\rho_{\rm{eff}}.
\end{align}}
Therefore, when considering the single-photon loss and pure dephasing, the only remaining error in the computational subspace is the bit flip characterized by the operator  
$\sigma_{n}^{x}$, which commutes with the evolution operator $U_{\rm{MS}}(t)$.
Therefore, an erroneous gate operation is equivalent to an error-free gate followed 
by an error $\sigma_{n}^{x}$. Therefore, our cat-code gates preserve the error bias.

\end{widetext}

{
To be specific, we can assume that the dominant error $\sigma_{n}^{x}$ occurs in
one of the cat-state qubits at time $\tau_{\rm{err}}$ ($0<\tau_{\rm{err}}<t_{g}$). Then, the evolution should be modified as
\begin{align}
	U_{\rm{MS}}^{\rm{err}}(t_{g})=&U_{\rm{MS}}(t_{g}-\tau_{\rm{err}})\sigma_{x}^{n}U_{\rm{MS}}(\tau_{\rm{err}}).                        
\end{align} 
As shown in our manuscript, the evolution operator $U_{\rm{MS}}(t)$ reads 
\begin{align}
	U_{\rm{MS}}(t)=\exp{\left[-i\chi(t)a_{0}^{\dag}S_{x}+{\rm{h.c.}}\right]}\exp{\left[-i\beta(t)S_{x}^{2}\right]},
\end{align}
where $S_{x}=\frac{1}{2}\sum_{n}\sigma_{n}^{x}$ commutes with the dominant error $\sigma_{n}^{x}$.
Therefore,
we obtain 
\begin{align}
	U_{\rm{MS}}^{\rm{err}}(t_{g})=\sigma_{x}^{n}U_{\rm{MS}}(t_{g}),
\end{align}
which indicates that our cat-code MS gate preserves the error bias.
}

\begin{figure}
	\centering
	\scalebox{0.4}{\includegraphics{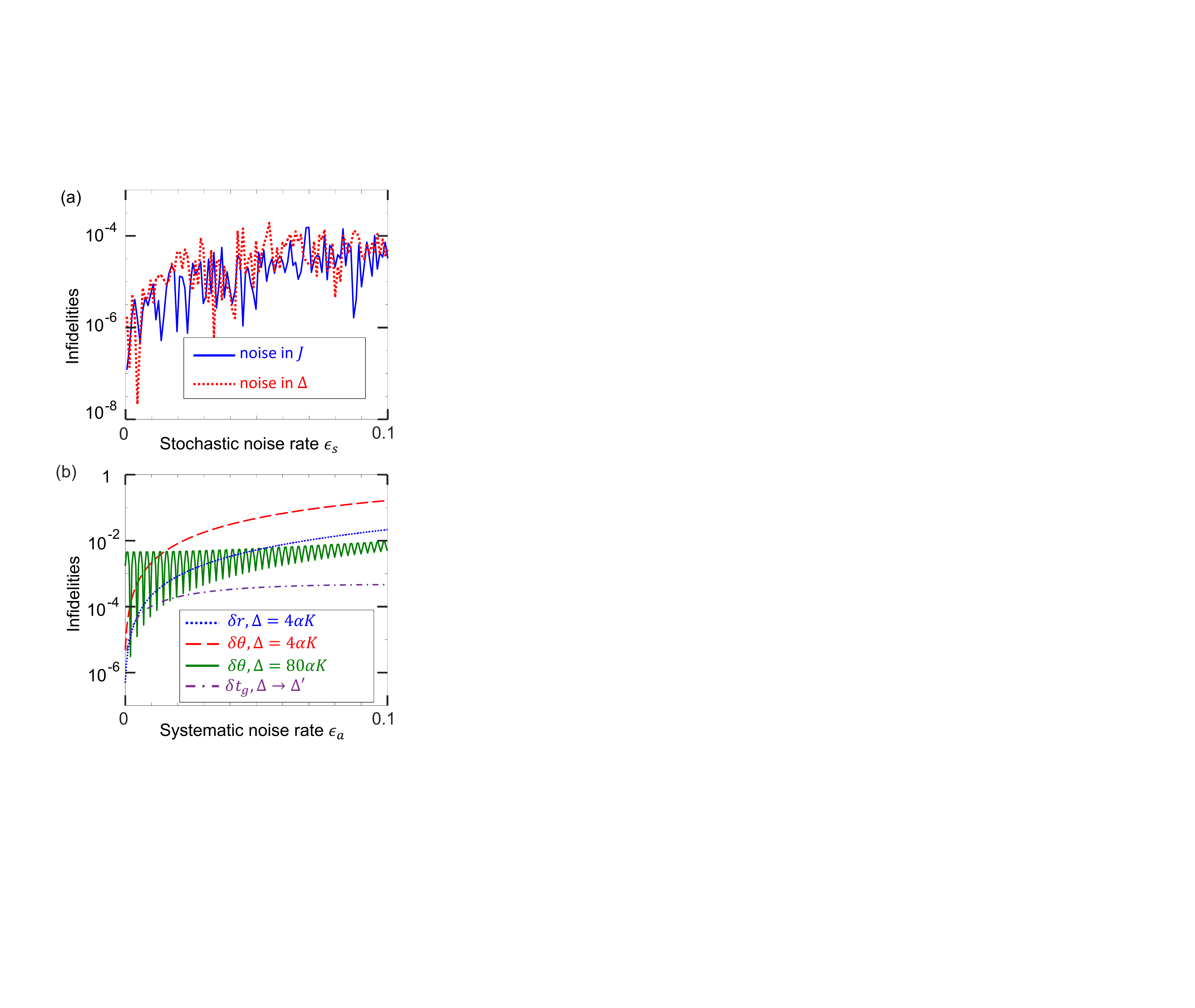}}
	\caption{{Noise-induced infidelities of the two-qubit gate when $\alpha=2$, {calculated for the total Hamiltonian $H$ in the presence of parameter imperfections}. (a) Infidelities $|\bar{F}_{2}(\epsilon_{s})-\bar{F}_{2}(0)|$ 
			versus stochastic noise rate $\epsilon_{s}$. (b) Infidelities $|\bar{F}_{2}(\epsilon_{a})-\bar{F}_{2}(0)|$ versus systematic noise rate $\epsilon_{a}$. 
			Here, the gate fidelity $\bar{F}_{2}$ is defined in Eq.~(\ref{eq17}).
			To identify the influence of parameter imperfections, we assume decay rates $\gamma=\kappa=0$.
			For (a), we choose
			$J/2\pi=K/2\pi=5~{\rm{MHz}}$ and $\Delta=4J\alpha$ ($m=1$), resulting in a gate time $t_{g}=25~{\rm{ns}}$. 
			For (b), when discussing a kind of noise, e.g., $\delta r$, we ignore the other noises.		   
			The purple dash-dot curve is calculated by changing the detuning $\Delta/2\pi=\sqrt{0.95}\times40~{\rm{MHz}}$ to $\Delta'/2\pi=\sqrt{0.05}\times40~{\rm{MHz}}$ at time $\tau=0.95t_{g}$.
			}
	}
	\label{fig3}
\end{figure}

However, pure dephasing in the KPOs causes transitions to the excited eigenstates $|\psi_{\pm}^{e,1}\rangle_{n}$ [the last term in Eq.~(\ref{eq6})] \cite{PuriPrx2019}.
Such transitions cause an infidelity ($1-F_{\rm{out}}$) that is equivalent to the leakage probability ($1-P_{\mathcal{C}}$).
This is demonstrated in Fig.~\ref{fig2}(c) that $P_{\mathcal{C}}\simeq F_{\rm{out}}$ in the presence of only pure
dephasing in the KPOs. Hence, in experiments to realize our protocol, it would be better
to choose systems with small dephasing rates.

%Engineering a strong two-photon dissipation channel with a dissipation rate $\kappa_{\rm{2ph}}$ for each
%KPOs can correct the leakage at a rate with a rate $$

{\section{Parameter imperfections}

In addition to decoherence, parameter imperfections may also cause 
infidelities. In the presence of parameter imperfections, 
a parameter $*$ should be corrected as $*'=*\pm\delta *$,
where $\delta *$ denotes the noise. For clarity, the noise-disturbed gate fidelity
is expressed as $\bar{F}_{N}(\delta *)$.
We consider two kinds of noise: stochastic and systematic.
For the stochastic noise, the noise rate $(\delta *)/*$ is a time-dependent random number; and can
be expressed as a random number $(\delta *)/*=\rm{rand}(\epsilon_{s})$ in the interval $(-\epsilon_{s},\epsilon_{s})$.
For instance, we consider stochastic noise in the parameters $J$ and $\Delta$.
The actual values of $J$ and $\Delta$ should be corrected as
\begin{align}
  J\left(t_{\eta}\right)=&J\left[1\pm {\rm{rand}}(\epsilon_{s})\right],\cr  \Delta\left(t_{\eta}\right)=&\Delta\left[1\pm {\rm{rand}}(\epsilon_{s})\right].
\end{align}
Here, $t_{\eta}$ means that, at time $0<t_{\eta}<t_{g}$, noise arises for the $\eta$th time.
Assuming that the noise randomly arises a total of $1,000$ times, the noise-induced infidelities $\left|\bar{F}_{N}(\epsilon_{s})-\bar{F}(0)\right|$
are very small, as shown in Fig.~\ref{fig3}(a). A noise rate $\epsilon_{s}=0.1$ only causes
an infidelity $\sim 10^{-4}$, indicating that the gates are mostly insensitive to stochastic noise.
The oscillations in the gate infidelities demonstrate that the stochastic noise randomly affects the system.

{
	\renewcommand\arraystretch{1.3}
	\begin{table*}
		\centering
		\caption{Fidelities of quantum gates based on bosonic codes. Coherence properties: Energy relaxation
			time ($T_{1}=1/\kappa_{j}$) and dephasing time ($T_{2}^{*}=1/\gamma_j$).}
		\label{tab1}
		\begin{tabular}{lllllll}
			\hline
			\hline
			Year & \ \ \ \ Code &\ \  \ \ \ \ Gate Type &\ \ \ \ $T_{1}$ ($\upmu{\rm{s}}$) &\ \ \ \ $T_{2}^{*}$ ($\upmu{\rm{s}}$) & \ \ \ \ Fidelity ($\%$)   \\
			\hline
			2017 \cite{Heeres2017} &\ \  \ \ Cat &\ \ \ \ \ \  Single-qubit gates &\ \ \ \ $\sim 170$ &\  \ \ \ $\sim43$&\ \ \ \ $98.5$ \\
			2018 \cite{Rosenblum2018} &\ \ \ \ Binomial &\ \ \ \ \ \ Controlled Not (CNOT) &\ \ \ \ $\sim2000$&\ \ \ \ $\sim500$&\ \ \ \ $89.0$ \\
			2018 \cite{Chou2018} &\ \ \ \ Binomial &\ \ \ \ \ \ Teleported CNOT &\ \ \ \ $\sim 1000$ &\ \ \ \ $\sim 300$&\ \ \ \  $79.0$ \\			
			2019 \cite{Hu2019} &\ \ \ \ Binomial &\ \ \ \ \ \ Single-qubit gates &\ \ \ \ $\sim140$ &\ \ \ \  $\sim 250$ &\ \ \ \ $97.0$ \\
			2019 \cite{Gao2019} &\ \ \ \ Fock  &\ \ \ \ \ \ eSWAP &\ \ \ \ $\gtrsim200$&\ \ \ \ $\gtrsim 300$&\ \ \ \ $85.0$ \\
			2020 \cite{Xu2020} &\ \ \ \ Binomial \& Cat &\ \ \ \ \ \ Geometric cPhase &\ \ \  \ $\gtrsim 500$&\ \ \ \ $\gtrsim300$&\ \ \ \ $89.4$\\
			\hline
			\multirow{3}*{Our protocol} & \ \ \ \ \multirow{3}*{Cat}  &\ \ \ \ \ Two-qubit MS gate &\ \ \ \ \multirow{3}*{$\sim 200$} &\ \ \ \ \multirow{3}*{$\sim 200$}&\ \ \ \ $\sim98$ \\
			~ & ~&\ \ \ \ \ Three-qubit MS gate &~ &~ & \ \ \  \ $\sim97$ \\
			~ & ~&\ \ \ \ \ Four-qubit MS gate &~ &~ &\ \  \ \ $\sim90$ \\
			\hline
			\hline
		\end{tabular}
	\end{table*}
}

For the systematic noise, the noise rate $(\delta *)/*=\epsilon_{a}$ becomes a small constant.
According to the evolution operator $U_{\rm{MS}}(t)$,
parameter imperfections may induce deviations in
the radius $r$ and the rotation angle $\theta$ that cause infidelities.
For simplicity, we can analyze 
the influence of imperfections in $r$ (caused by imperfections in $J$, $\alpha$, or $1/\Delta$) and $\theta$ (caused by imperfections in $\Delta$ or $t_{g}$).
As shown in Fig.~\ref{fig3}(b), the imperfections in $\theta$ (red-dashed curve) have a greater
influence than those of $r$ (blue-dotted curve) when fixing the detuning $\Delta$. This is because $\delta \theta$ can cause
excitations in the cavity mode $a_{0}$ [i.e., $\chi(t_{g})\neq 0$ in $U_{\rm{MS}}(t_{g})$], leading to infidelities.
These excitations can be suppressed by increasing the detuning $\Delta$ [see the green-solid curve in Fig.~\ref{fig3}(b)], 
because $\chi(t)$
is inversely proportional to $\Delta$.

However, a larger detuning means a longer gate time, which increases the influence of decoherence.
Note that imperfections in $\theta$ are mainly caused by imperfections in the gate time $t_{g}$,
which affect the system in the time interval $t_{g}\left(1-\epsilon_{a},~1+\epsilon_{a}\right)$.
We can increase $\Delta$ only in this time interval to minimize the influence on the gate time. 
To this end, we choose 
\begin{align}
   \Delta=\frac{4\sqrt{m}J\alpha}{\sqrt{1-\epsilon_{a}}}\ \ \ \  {\rm{and}} \ \ \ \
    \tau=t_{g}(1-\epsilon_{a})=\frac{2\pi m}{\Delta},
\end{align}
to satisfy $\chi(\tau)=0$.
Then, the detuning is increased to $\Delta'=4\sqrt{m'}J\alpha/\sqrt{\epsilon_{a}}$, where
$m'$ denotes the number of evolution cycles in phase space in the time interval $t_{g}\left(1-\epsilon_{a},~1\right)$.
These parameters ensure that the total geometric phase is still $\beta(t_{g})=-\pi/2$.
The gate time becomes
\begin{align*}
t_{g}=\frac{\pi}{2J\alpha}\left[\sqrt{m(1-\epsilon_{a})}+\sqrt{m'\epsilon_{a}}\right]\approx \frac{\sqrt{m}\pi}{2J\alpha},
\end{align*}
for $m\geq m'$ and $\epsilon_{a}\ll 1$. Therefore, the gate time is mostly unchanged, while we can achieve the gate robustness against its parameter imperfections [see the purple dash-dot curve in Fig.~\ref{fig3}(b)]. }

%For $\alpha^2\gtrsim 3$, the bit-flip error rate $\kappa\alpha^2$ increases 
%linearly with the photon number, which becomes a kind of ``biased noise'' \cite{Aliferis2008}. 
%Such noise is an important resource in fault-tolerant quantum computation \cite{Aliferis2008,Tuckett2018}, because
%it can reduce the number of building blocks for error-correction codes by 
%focusing strongly on the remaining bit-flip error \cite{PuriPrx2019}.

\begin{figure}
	\centering
	\scalebox{0.42}{\includegraphics{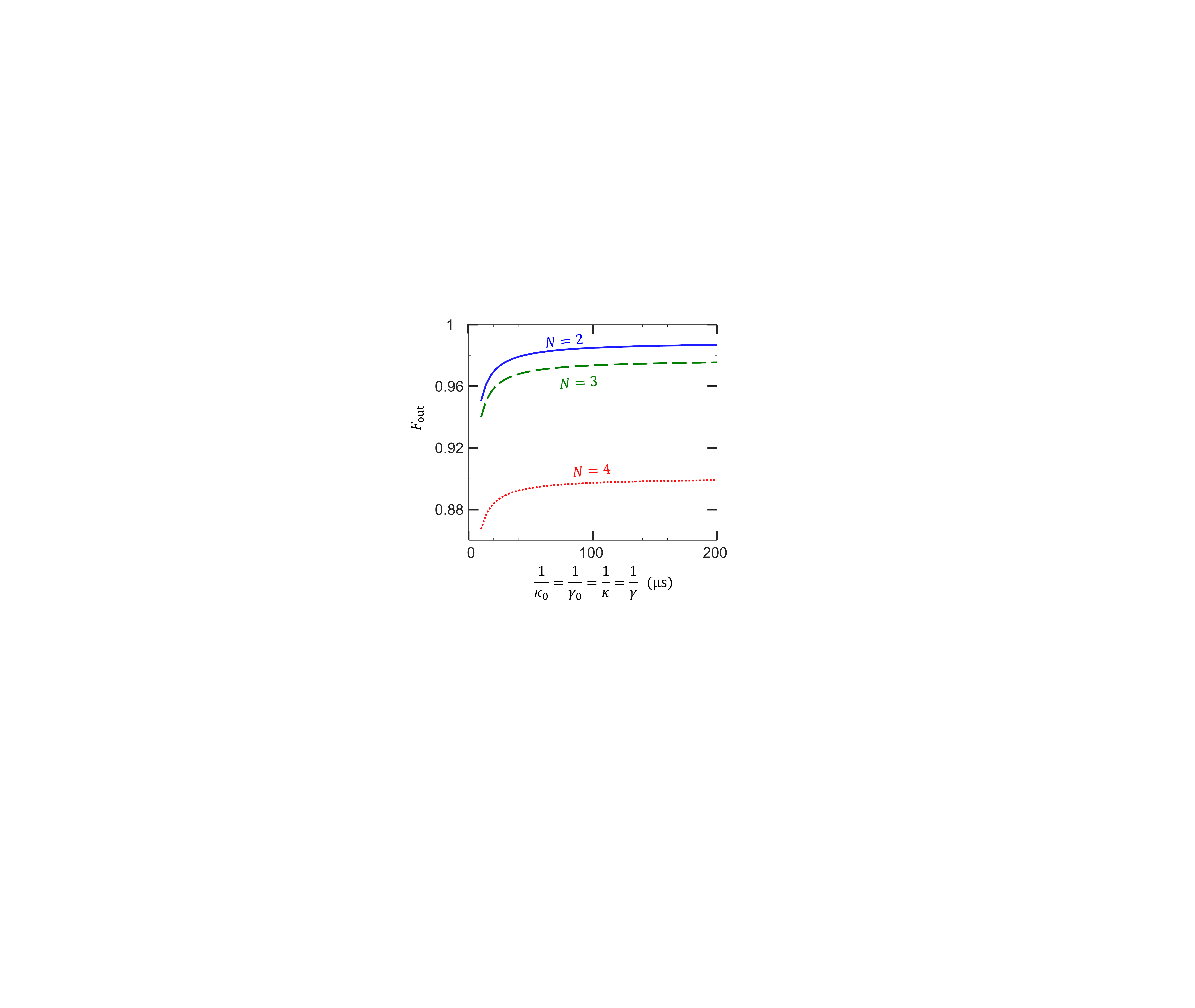}}
	\caption{{Output-state fidelities $F_{\rm{out}}$ of the $N$-qubit gates in the presence of 
			decoherence and parameter imperfections. {Here, $F_{\rm{out}}$ is calculated using the Lindblad master equation Eq.~(\ref{eq4a}) for the total Hamiltonian $H$ in the presence of parameter imperfections.}
			Parameters are $J/2\pi=K/2\pi=5~{\rm{MHz}}$ and $\alpha=2$.
			We choose systematic noise rates 
			$\delta J/J=\delta \Delta/\Delta=\delta\Delta'/\Delta'=\delta t_{g}/t_{g}=-\epsilon_{a}=-5\%$.
			We change the detuning $\Delta/2\pi=\sqrt{0.95}\times40~{\rm{MHz}}$ to $\Delta'/2\pi=\sqrt{0.05}\times40~{\rm{MHz}}$ at the time $\tau=0.95t_{g}$, to suppress the error induced by parameter imperfections in the gate time.}
	}
	\label{fig3b}
\end{figure}

\section{Discussion}
{Using the above optimized method, when decoherence and parameter imperfections are considered,
the fidelities of the output states for $N=2,3,4$ are shown in Fig.~\ref{fig3b}.
The rates of the systematic noise are chosen as
\begin{align}
   \delta J/J=\delta \Delta/\Delta=\delta\Delta'/\Delta'=\delta t_{g}/t_{g}=-5\%.
\end{align}
We ignore the stochastic noise because, practically, it does not affect the system dynamics.}
Superconducting circuits \cite{Koch2007,You2007,Flurin2015,Grimm2020,Wustmann2013,Gu2017,Krantz2019APR,Kjaergaard2020,Kwon2021Jap} can be a possible implementation of our protocol (see the details in Appendix~\ref{App3}). 
For instance, one can use the Josephson parametric amplifier \cite{Kockum2019,Nation2012,Wallquist2006,Liao2010,Xiang2013,Yaakobi2013,Macklin2015,Roy2016,Gu2017,Wang2019,WangPrx2019,Kjaergaard2020,Masuda2021} to realize the Hamiltonian $H_{n}^{\rm{Kerr}}$.
Another especially promising setup to realize our protocol could be a
single junction or transmon embedded in a 3D oscillator \cite{Cai2021,Ma2021}. The
Kerr nonlinearity and the two-photon drive can be respectively realized
by the Josephson junction (transmon) nonlinearity and four-wave mixing \cite{Chen2019,Qin2019,Qin2020,Qin2021,PuriPrx2019}.
The change of detuning can be generally realized by changing the frequency $\omega_{c}$ (see Appendix~\ref{App4} for more details).
Such a change should be as fast as possible to avoid 
introducing an additional phase shift.

{Note that the cat-state qubits discussed in our manuscript belong to a larger family of bosonic qubits.}
{Bosonic-code quantum gates have been realized using superconducting circuit quantum electrodynamics (circuit QED) architecture
and three-dimensional (3D) cavities, especially 3D coaxial cavities.
{The experimental platforms that have already realized different bosonic qubits could implement bosonic cat-state qubits \cite{Cai2021,Ma2021}. For instance,
Ref.~\cite{Xu2020} reported an experimental realization of both binomial and cat-state qubits using the same experimental platform.}
For clarity, we show the fidelities and the corresponding coherence properties of some one- and two-qubit gates in Table \ref{tab1}, which have been realized in current experiments.
As shown, it is still challenging to achieve high-fidelity bosonic gates in current experiments, which 
may lower the code-capacity threshold for error correction.

{Geometric quantum gates with cat-state qubits were recently experimentally realized in 2020 \cite{Xu2020}.
With coherence times $T_{1}=1/\kappa_{j}\sim500$~$\upmu$s and $T_{2}^{*}=1/\gamma_{j}\sim300$~$\upmu$s, that experiment \cite{Xu2020} only realized two-qubit gates with fidelities
$\sim 90\%$.}
{In contrast to this, our protocol can easily generate a two-qubit gate with fidelities $\gtrsim 90\%$ even when using much shorter coherence times [blue-solid curve in 
Fig.~\ref{fig3b}].
For the implementation shown in Appendix \ref{App3}, the experimental coherence times for the Kerr parametric oscillator can reach $T_{1}\sim T_{2}^{*}\gtrsim15 \mu{\rm s}$ \cite{Wang2019}, which enable our protocol to generate two-, three-, and four-qubit gates with fidelities $\sim 97.5\%$, $\sim94\%$, and $\sim 86\%$, respectively.}

\section{Conclusions}
We investigate the possibility
of using photonic cat-state qubits for implementing multiqubit geometric gates,
which can generate maximally multimode entangled cat states with high fidelities.
%The gates 
%can rapidly transform product states into maximally multi-mode entangled cat states with high fidelities.
%We have analyzed the influence of various types of noise on the protocol. 
Our theoretical protocol is robust against stochastic noise along the evolution path because of the character of the geometric evolution. 
By increasing the detuning at a suitable time, the protocol can tolerate imperfections
in the gate time.
For large $\alpha$, the phase-flip error can be exponentially suppressed, leaving 
only the bit-flip error. %, which is caused by the single-photon loss of the cavity modes.
The pure dephasing of the cavity modes may lead to photon leakage out of the computing subspace, 
but does not cause qubit-dephasing problems for the system.
{This dominant error commutes with the evolution operator, 
which means that our MS gates preserve the error bias.}
%Therefore, our protocol based on cat-state qubits can reduce error channels in the computing subspace. Thus,
Therefore, error-correction layers can focus only on the bit-flip error that uses less physical resources.
In summary, our results offer a
realistic and hardware-efficient method for multiqubit fault-tolerant
quantum computation.

%For instance, Fig.~\ref{fig2a}(a) exactly shows the quantum phenomenon of quantum state collapse and revival,
%which manifests most clearly for a vanishing qubit term in ultrastrong light-matter coupling \cite{}.
%In this manuscript, we focus on the cat-code geometric gates.

\appendix

\section{Arbitrary single-qubit rotations of cat-state qubits}\label{App2}
	
\begin{figure*}
 \centering
	\scalebox{0.6}{\includegraphics{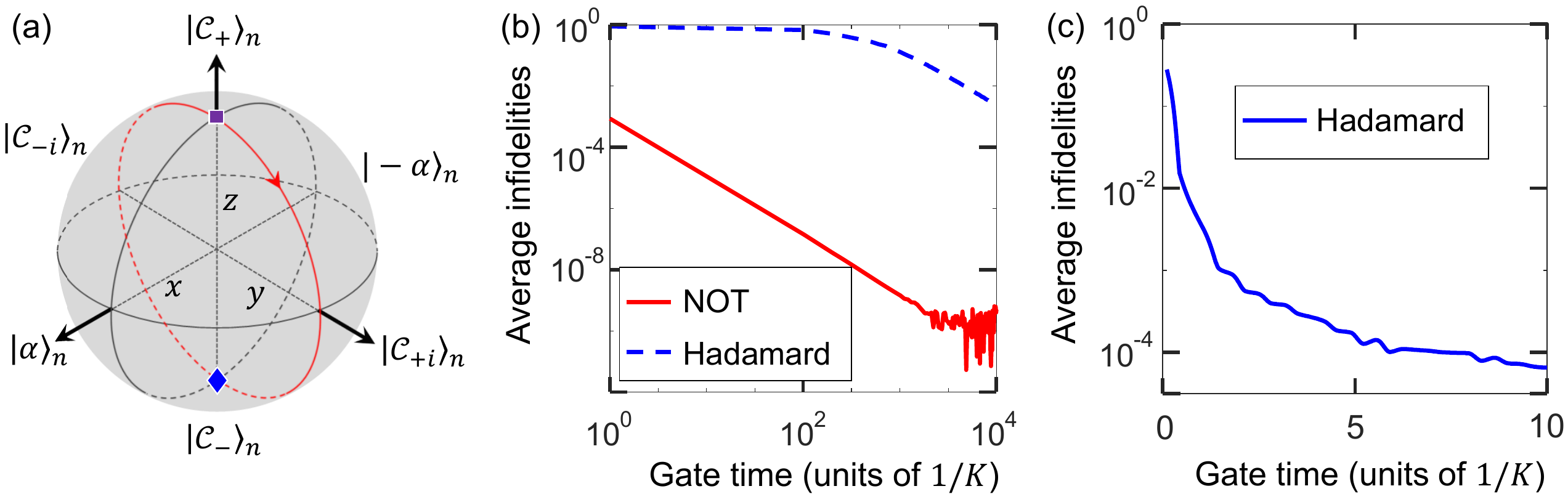}}
	\caption{(a) Bloch sphere of the cat qubit in the limit of large $\alpha$ (i.e., $\alpha=2$). The red circle with a red arrow denotes 
		the evolution path for the NOT gate. For instance, when the input state is $|\mathcal{C}_{+}\rangle_{n}$ (purple square), the NOT gate transforms this input state into $|\mathcal{C}_{-}\rangle_{n}$ (blue diamond). The states on all $y$ axis are
		$|\mathcal{C}_{\pm i}\rangle_{n}\simeq (|\alpha\rangle_{n}\pm i|-\alpha\rangle_{n})/\sqrt{2}$.  
		(b) Average infidelities of the Hadamard and NOT gates versus the gate time calculated via the Hamiltonian in Eq.~(\ref{eqS9}). (c) Average infidelities of the Hadamard gate when the additional Hamiltonian in Eq.~(\ref{eqS12}) is added, i.e., when the total Hamiltonian is ${H}_{\rm{tot}}(t)=\tilde{H}_{n}^{\rm{Kerr}}+H_{\rm{add}}(t)$. We assume that
		the frequency of each KPO is $\omega_{c}=800K$ and the coherent-state amplitude is $\alpha=2$. Other parameters are given below Eq.~(\ref{eqS11}). 
	}
	\label{figS2}
\end{figure*}	
	
	Accompanied by a variety of single-qubit rotations, the M{\o}lmer-S{\o}rensen gate can be adapted to many quantum
	algorithms, such as Grover's quantum search algorithm~\cite{Brickman2005,Grover1997,Nielsen2000}.
	To realize such single-qubit rotations, one needs to add a single-photon drive to each KPO~\cite{Grimm2020}. The Hamiltonian 
	for each KPO becomes
	\begin{align}\label{eqS9}
		\tilde{H}_{n}^{\rm{Kerr}}=&\Omega_{p}\left(a_{n}^{\dag 2}+a_{n}^{2}\right)-K a_{n}^{\dag 2}a_{n}^{2}\cr
		&+\Delta_{q}a_{n}^{\dag}a_{n}+(\xi_{p}a_{n}+\xi_{p}^{*}a_{n}^{\dag}),
	\end{align}
	where $\xi_{p}$ is the complex driving amplitude.
	Note that the parameters discussed in this section are independent of those in the main text and Appendix B.
	When 
	\begin{align} 
		\Delta_{q},|\xi_{p}|\ll E_{\rm{gap}},
	\end{align}
 the evolution is restricted in the cat-state subspace $\mathcal{C}$.
	The effective Hamiltonian in the cat-subspace reads ($\alpha=\alpha^{*}=\sqrt{\Omega_{p}/K}$):
	\begin{align}\label{eqS10}
		\tilde{H}_{n,\rm{eff}}^{\rm{Kerr}}=&\frac{1}{2}\Delta_{q}\alpha^{2}\left(\coth \alpha^2-\tanh\alpha^{2}\right)\sigma_{n}^{z}\cr&+\left[\left(\xi\alpha\sqrt{\tanh\alpha^{2}}+
		\xi^{*}\alpha\sqrt{\coth\alpha^{2}}\right)\sigma_{n}^{-}+{\rm{h.c.}}\right]\cr
		=&\frac{\tilde{\Delta}_{q}}{2}\sigma_{n}^{z}+\left[\Omega_{1}\exp{(-i\varphi)}\sigma_{n}^{-}+{\rm{h.c.}}\right],
	\end{align}
	where $\sigma_{n}^{z}=|\mathcal{C}_{-}\rangle_{n}\langle\mathcal{C}_{-}|-|\mathcal{C}_{+}\rangle_{n}\langle\mathcal{C}_{+}|$.

	Obviously, the effective Hamiltonian $\tilde{H}_{n,{\rm{eff}}}^{\rm{Kerr}}$ contains all the Pauli matrixes for a two-level system. Thus, it can realize 
	arbitrary single-qubit rotations. The evolution operator in matrix form becomes 
	\begin{widetext}
	\begin{align*}\label{eqS11}
		U_{1}=\exp{\left(-i\tilde{H}_{n,{\rm{eff}}}t\right)}=\left(\begin{array}{cc}
			\cos(\Xi t)-i\sin(\Xi t)\cos\theta & -i\exp{(-i\varphi)}\sin(\Xi t)\sin\theta\cr
			-i\exp{(i\varphi)}\sin(\Xi t)\sin\theta & 
			\cos(\Xi t)+i\sin(\Xi t)\cos\theta
		\end{array}\right),
	\end{align*}
	\end{widetext}
	which denotes an arbitrary rotation on the Bloch sphere [see Fig.~\ref{figS2}(a)]. 
	Here, 
	\begin{align}
	  \Xi&=\sqrt{{\tilde{\Delta}_{q}^{2}/4}+\Omega_{1}^{2}},\cr
	  \theta&=\arctan(2\Omega_{1}/\tilde{\Delta}_{q}).
	\end{align}
	For instance, when $\Xi t=\pi/2$, $\theta=\pi/4$, and $\varphi=0$, 
	$U_{1}$ denotes the Hadamard gate up to a global phase $\pi/2$ [see the blue-dashed curve in Fig.~\ref{figS2}(b)]. When $\Xi t=\pi/2$, $\theta=\pi/2$, and $\varphi=0$, $U_{1}$ 
	becomes the NOT gate up to a global phase $\pi/2$ [see the red-solid curve in Fig.~\ref{figS2}(b)].
	We can see in Fig.~\ref{figS2}(b) that the gate time of the Hadamard gate is much longer than that of the NOT gate.
	This is understood because the effective detuning $\tilde{\Delta}_{q}$ exponentially decreases when $\alpha$ increases.
	Thus, it takes a long time to obtain a phase rotation about
	the $z$ axis.

	As an alternative to obtaining a large effective detuning $\tilde{\Delta}_{q}$, one can employ an interaction Hamiltonian 
	\begin{align}\label{eqS12}
		H_{\rm{add}}(t)=&\xi_{J}\cos\left[\varphi_{a}(a_{n}e^{-i\omega_{c}t}+a_{n}^{\dag}e^{i\omega_{c}t})\right]\cr
		               =&\frac{\xi_{J}}{2}\left[D_{n}(\beta_{t})+D_{n}(-\beta_{t})\right],
	\end{align}
	which can be realized by strongly coupling a high impedance cavity mode to a Josephson junction~\cite{Masluk2012,Pop2014,Cohen2017}. 
	Here, $\xi_{J}$ is the effective Josephson energy and $\varphi_{a}=\sqrt{Z_{a}/2R_{Q}}$, with $Z_{a}$ and
	$R_{Q}$ being the impedance
	of the cavity mode seen by the junction and the superconducting resistance
	quantum, respectively.
	The displacement parameter is 
	\begin{align}
		\beta_{t}=i\varphi_{a}\exp{(i\omega_{c}t)}.
	\end{align}
	When $\omega_{c},E_{\rm{gap}}\gg \xi_{J}$ and $\varphi_{a}\simeq 2\alpha$, the effective Hamiltonian under the rotating wave approximation in the cat-state subspace becomes \cite{Cohen2017}
	\begin{align}\label{eqS13}
		\tilde{H}_{\rm{add}}=\frac{\tilde{\Delta}_{q}}{2}\sigma_{n}^{z},
	\end{align}
	where $\tilde{\Delta}_{q}\simeq \xi_{J}/\alpha\sqrt{2\pi}$. 
	Substituting Eq.~(\ref{eqS13}) into Eq.~(\ref{eqS10}) and assuming $\Delta_{q}=0$, 
	the evolution operator still takes the form of Eq.~(\ref{eqS11}). 
	Figure \ref{figS2}(c) shows the average infidelities of the Hadamard gate when the additional Hamiltonian $H_{\rm{add}}(t)$ is added.
	Comparing to the result in Fig.~\ref{figS2}(b), the additional Hamiltonian $H_{\rm{add}}(t)$
	obviously increases the effective detuning, so that the gate time is shortened.
	For instance, a gate time $\sim 5/K\approx 8~{\rm{ns}}$ (for $K/2\pi\sim 10~{\rm{MHz}}$) is enough 
	to achieve a Hadamard gate with a fidelity $\gtrsim 99.99\%$.

\section{Preparing Schr\"odinger cat states}\label{App1}
To generate the quantum cat states in the KPOs, 
we first decouple the KPOs from the common cavity $a_{0}$ by tuning $J=0$ or $\Delta=\infty$.
Then, we change the Hamiltonian for each KPO
to be time-dependent [we assume $t\in[-t_{0},0]$ and  $\Omega_{p}(t)=\Omega_{p}^{*}(t)\geq0$ for simplicity]:
\begin{align}\label{eqA1}
	H_{n}^{\rm{Kerr}}(t)=&\Omega_{p}(t)\left(a_{n}^{\dag 2}+a_{n}^{2}\right)-K a_{n}^{\dag 2}a_{n}^{2}\cr
	&+\Delta_{q}(t)a_{n}^{\dag}a_{n},
\end{align}
where $\Delta_{q}(t)=\omega_{c}-\omega_{p}/2$ is a time-dependent detuning and $t_{0}$ denotes the total evolution time required for the generation of cat states. 
To study the dynamics of the time-dependent Hamiltonian $H_{n}^{\rm{Kerr}}(t)$, we introduce
the displacement operators $D_{n}(\pm \alpha_{t})=\exp{\left(\pm\alpha_{t} a_{n}^{\dag}\mp \alpha_{t} a_{n}\right)}$
to transform $H_{n}^{\rm{Kerr}}(t)$ as
\begin{align}
	H'_{n}(t)=&D_{n}(\pm\alpha_{t})H_{n}^{\rm{Kerr}}(t)D_{n}(\mp\alpha_{t})-iD_{n}(\pm\alpha_{t})\dot{D}_{n}(\mp\alpha_{t})\cr\cr
	=&\left[\Delta_{q}(t)-4K\alpha_{t}^{2}\right]a_{n}^{\dag}a_{n}\mp 2K\alpha_{t}\left(a_{n}^{\dag 2}a_{n}+a_{n}^{\dag}a_{n}^{2}\right)\cr
	&-Ka_{n}^{\dag 2}a_{n}^{2}\mp\left[\alpha_{t}\Delta_{q}(t)+i\dot{\alpha}_{t}\right]a_{n}^{\dag}\cr
	&\mp\left[\alpha_{t}\Delta_{q}(t)-i\dot{\alpha}_{t}\right]a_{n},
\end{align}
where $\alpha_{t}=\sqrt{\Omega_{p}(t)/K}\geq 0$ is the time-dependent amplitude of a coherent state $|\alpha_{t}\rangle$.

Obviously, when 
\begin{align}\label{eqS6}
	\left[\Delta_{q}(t)-4K\alpha_{t}^{2}\right]~&\gg~2K\alpha_{t},\cr\cr 	\left[\Delta_{q}(t)-4K\alpha_{t}^{2}\right]~&\gg\sqrt{\left[\alpha_{t}\Delta_{q}(t)\right]^{2}+\dot{\alpha}_{t}^{2}},
\end{align}
the Hamiltonian $H'_{n}(t)$ cannot change the photon number of the system in the displacement frame.
In this case, 
when $\alpha_{t}$ satisfies the boundaries
\begin{align}
	\alpha_{t}|_{t=-t_{0}}=0, \ \ \ {\rm{and}} \ \ \ \alpha_{t}|_{t=0}=\alpha.
\end{align}
Assuming that the system in the displaced frame is in the displaced vacuum state $|0\rangle_{n}$ at the time $-t_{0}$, the evolution in the lab frame
can be described by
\begin{align} 
	|\psi(t)\rangle_{n}=D_{n}(\pm\alpha_{t})|0\rangle_{n},
\end{align}
or can be equivalently described by
\begin{align}
	|\psi(t)\rangle_{n}=\mathcal{N}_{\pm}(\alpha_{t})\left[D_{n}(\alpha_{t})\pm D_{n}(-\alpha_{t})\right]|0\rangle_{n},
\end{align}
where $\mathcal{N}_{\pm}(\alpha_{t})=1/\sqrt{2\left[1\pm\exp(-2\alpha_{t}^{2})\right]}$.

\begin{figure*}
	\centering
	\scalebox{0.37}{\includegraphics{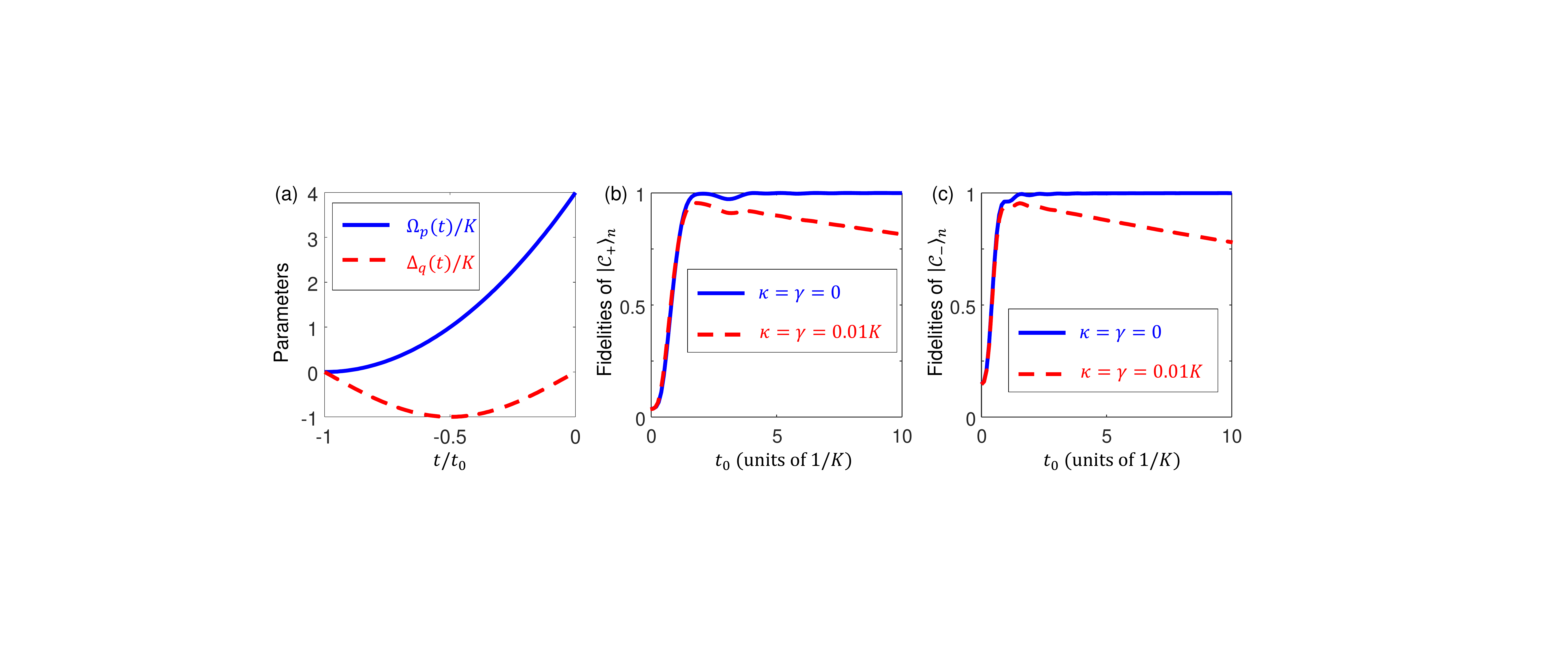}}
	\caption{(a) Parameters used for the generation of the cat states $|\mathcal{C}_{\pm}\rangle_{n}$. Fidelities of (b) the even cat state $|\mathcal{C}_{+}\rangle_{n}$ and (c) the odd cat state $|\mathcal{C}_{-}\rangle_{n}$ versus the total evolution time $t_{0}$ calculated for the Hamiltonian in Eq.~(\ref{eqA1}). The initial states for (b) is $|\psi(-t_{0})\rangle_{n}=|0\rangle_{n}$ and (c) is $|\psi(-t_{0})\rangle_{n}=|1\rangle_{n}$. Other parameters are given in Eq.~(\ref{eqS7}).
	}
	\label{figS1}
\end{figure*}
To satisfy the condition in Eq.~(\ref{eqS6}), for $t\leq0$, we assume $\alpha_{t}=\alpha$ and $\Delta_{q}(t)=0$, while for $\left(t_{0}\leq t<0\right)$
we assume 
\begin{align}\label{eqS7}
	\alpha_{t}&=\frac{\alpha}{t_{0}} (t+t_{0}), \cr
	\Delta_{q}(t)&=-K\sin\left[\frac{\pi}{t_{0}}(t+t_{0})\right].
\end{align}
Then, at $t=0$, the desired cat states $|\mathcal{C}_{\pm}\rangle_{n}=|\psi(0)\rangle_{n}$ can be generated. 
The driving amplitude $\Omega_{p}(t)$ and the detuning $\Delta_{q}(t)$ using the parameters in Eq.~(\ref{eqS7}) 
are shown in Fig.~\ref{figS1}(a). In the absence of decoherence, the fidelities 
\begin{align}
	F_{\pm}={}_{n}\langle \mathcal{C}_{\pm}|\rho(0)|\mathcal{C}_{\pm}\rangle_{n},
\end{align}
of the prepared cat states are shown in Fig.~\ref{figS1}(b) and Fig~\ref{figS1}(c).
As a result, an evolution time $t_{0}\gtrsim 1.7/K\approx 3~{\rm{ns}}$ (when $K/2\pi=10~{\rm{MHz}}$) is enough to generate the cat states
$|\mathcal{C}_{\pm}\rangle_{n}$ with fidelities $\gtrsim 99\%$. In the presence of decoherence, for the $n$th KPO,
the dynamics is described by the  Lindblad master equation 
\begin{align}
	\dot{\rho}_{n}=-i[H_{n}^{\rm{Kerr}}(t),\rho_{n}]+\kappa \mathcal{D}[a_{n}]\rho_{n}+\gamma \mathcal{D}[a_{n}^{\dag}a_{n}]\rho_{n},
\end{align}
where 
\begin{align}
	\mathcal{D}[o]\rho_{n}=o\rho_{n} o^{\dag}-\frac{1}{2}\left(o^{\dag}o\rho_{n}+\rho_{n} o^{\dag}o\right)
\end{align} is the 
Lindblad superoperator, $\kappa$ is the single-photon loss rate, and $\gamma$ is the pure dephasing rate.
In Fig.~\ref{figS1}(b) and Fig.~\ref{figS1}(c), we can see that the fidelities of the cat states can be higher than 
$95\%$ when the decay rates are $\kappa=\gamma=0.01K$.

		\section{A possible implementation using superconducting quantum interference devices}\label{App3}
		\begin{figure*}
			\centering
			\scalebox{0.5}{\includegraphics{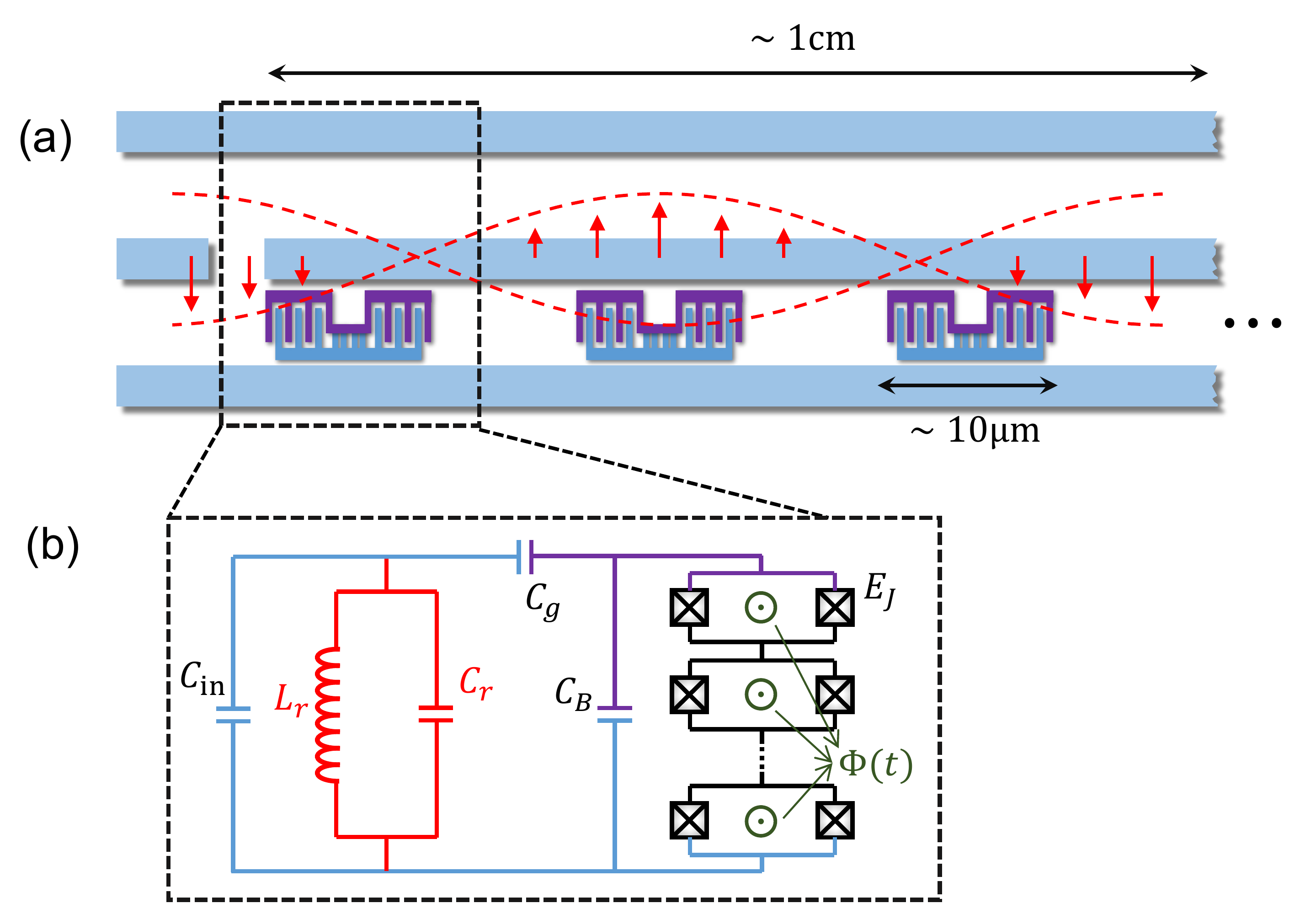}}
			\caption{(a) Simplified schematic of the transmon device design, which consists of several KPOs, shunted by a short section of a twin-lead transmission line. This short
				section of line can be well approximated as a lumped-element capacitor. (b) Effective circuit diagram
				of a KPO coupled to an $LC$ oscillator. The KPO is realized using an array of Josephson junctions, in which the Josephson energy $E_{J}$ is tunable by controlling the external magnetic flux $\Phi(t)$.
				The array of Josephson junctions with capacitance and Josephson
				energy $C_{J}$ and $E_{J}$ are shunted by an additional large capacitance
				$C_{B}$, matched by a comparably large gate capacitance $C_{g}$. {Following the standard quantization
				procedure for circuits \cite{Koch2007,You2007}, we have absorbed the junction capacitance $C_{J}$
				into the parallel capacitance $C_{B}$ for simplicity.}
			}
			\label{figS3}
		\end{figure*}
		
		A possible implementation for our protocol can be based on superconducting quantum interference devices (SQUIDs). 
		For instance, the KPOs can be realized using an array of Josephson junctions.
		Such quantum parametric oscillators have been experimentally realized, in e.g,
		Ref.~\cite{WangPrx2019}.
		We can then embed these parametric oscillators (with a relatively long distance to each other)
		in a transmission-line resonator [see Fig.~\ref{figS3}(a)].
		The transmission-line resonator can be modeled by an $LC$ oscillator [see Fig.~\ref{figS3}(b)]
		and it is used as the cavity mode $a_{0}$ in our protocol.
		The direct coupling between two adjacent KPOs can be neglected because of the long distance between them.

		Following the standard quantization
		procedure for circuits, 
		the Hamiltonian for the circuit in Fig.~\ref{figS3}(b)
		is
		\begin{align}\label{R32}
			H_{n}=&\frac{\hat{\phi}_{r}^{2}}{2L_{r}}+\frac{(C_{B}+C_{g})\hat{Q}_{r}^{2}}{2C_{*}}\cr
			&+\frac{(C_{g}+C_{\rm{in}}+C_{r})\hat{Q}_{J}^{2}}{2C_{*}}-N_{0}E_{J}[\Phi(t)]\cos\frac{\hat{\phi}}{N_{0}}\cr
			&+\frac{C_{g}\hat{Q}_{r}\hat{Q}_{J}}{C_{*}},
		\end{align}
	where
	    \begin{align}
			C_{*}=C_{B}C_{g}+C_{B}C_{\rm{in}}+C_{g}C_{\rm{in}}+C_{B}C_{r}+C_{g}C_{r}.
		\end{align}
		The subscript $n$ denotes that this is the Hamiltonian describing the coupling
		between the $n$th KPO and the cavity mode $a_{0}$.
		The first line in $H_{n}$ describes the local oscillator of the resonator $a_{0}$;
		the second line is the Hamiltonian for the KPO; and the third line describes the coupling.
		Here, $\hat{Q}_{r}$ and $\hat{Q}_{J}$ are charges for the $LC$ resonator and the array of Josephson junctions, respectively;
		$\hat{\phi}_{r}$ and $\Phi(t)$ are the branch and external-magnetic fluxes for modulating the energies of the quantum
		$LC$ circuit and the KPO, respectively; $N_{0}$ is the number of SQUIDs in the array; and $E_{J}$
		is the Josephson energy of a single SQUID.

		In the realistic limit of large resonator capacitance $C_{r}\gg \left( C_{B}+C_{g}\right)$,
		we can simplify the Hamiltonian $H_{n}$ as 
		\begin{align}
			H_{n}=&\omega_{0}a_{0}^{\dag}a_{0}+4E_{C}\hat{n}^2-N_{0}E_{J}[\Phi(t)]\cos\frac{\hat{\phi}}{N_{0}}\cr&+\frac{2C_{g}eV_{\rm{rms}}^{0}}{C_{g}+C_{B}} (a_{0}+a_{0}^{\dag})\hat{n}.
		\end{align}
		Here, $\hat{n}$ and $\hat{\phi}$ are the number of Cooper pairs and
		the overall phase across the junction array, respectively;
		$E_{C}$ is the KPO charging energy, and 
		$\omega_{0}=1/\sqrt{L_{r}C_{r}}$ denotes the frequency of the cavity mode $a_{0}$. Moreover,
		The
		root-mean-square voltage of the local oscillator is denoted by
		$V_{\rm{rms}}^{0}=\sqrt{\omega_{0}/2C_{r}}$.
		
		We assume that the Josephson energy $E_{J}$ is modified as (with a frequency $\omega_{p}$)
		\begin{align}
			E_{J}[\Phi(t)]=E_{J}+\delta E_{J}\cos(\omega_{p} t).
		\end{align}
		After applying the Taylor expansion of $\cos\left(\hat{\phi}/N_{0}\right)$ to
		fourth order, we obtain
		\begin{align}
			H_{n}\approx &~ \omega_{0}a_{0}^{\dag}a_{0}+4E_{C}\hat{n}^2-N_{0}E_{J}(1-\hat{X}+\hat{X}^2/6)
			\cr &- N_{0}\delta E_{J}(1-\hat{X})\cos(\omega_{p}t) \cr
			&+\frac{2C_{g}eV_{\rm{rms}}^{0}}{C_{g}+C_{B}} (a_{0}+a_{0}^{\dag})\hat{n},
		\end{align}
		where $\hat{X}=(\hat{\phi}/N_{0})^2/2$. We assume that the system is not highly excited, i.e., the highest level is much smaller than the dimension of the Hilbert space.
		Then, following the standard quantization
		procedure for circuits \cite{Koch2007,You2007}, we can define ($\hbar=1$)
		\begin{align}
			\hat{n}=&-i n_{0}(a_{n}-a_{n}^{\dag}),\cr  \hat{\phi}=&\phi_{0}(a_{n}+a_{n}^{\dag}),
		\end{align}
		where $n_{0}=\sqrt[4]{E_{J}/(32N_{0}E_{C})}$ and ${\phi_{0}}=2\sqrt{2}/n_0$ are the
		zero-point fluctuations.
		The quadratic time-independent part of the Hamiltonian $H_{n}$
		can be diagonalized and the Hamiltonian $H_{n}$ becomes 
		\begin{align}\label{R37}
			H_{n}=&~\omega_{0}a_{0}^{\dag}a_{0}+\omega_{c}a_{n}^{\dag}a_{n}-\frac{E_{C}}{12N_{0}^{2}}\left(a_{n}+a_{n}^{\dag}\right)^{4}
			\cr&+\frac{\delta E_{J}\omega_{c}}{4E_{J}}\left(a_{n}+a_{n}^{\dag}\right)^2 \cos(\omega_{p}t)\cr
			&+\frac{2C_{g}eV_{\rm{rms}}^{0}n_{0}}{C_{g}+C_{B}} (a_{0}+a_{0}^{\dag})(ia_{n}^{\dag}-ia_{n})
		\end{align}
		where $\omega_{c}=\sqrt{8E_{C}E_{J}/N_{0}}$. Here, we have dropped the constant terms for simplicity.
		
		We assume that the two-photon drive is resonant with the cavity mode, i.e., $2\omega_{p}=\omega_{c}$.
		When the conditions 
		\begin{align}
			\omega_{p}\gg &\frac{E_{C}}{12N_{0}^{2}},\cr   \omega_{p}\gg &\frac{\delta E_{J}\omega_{c}}{4E_{J}}, \cr \omega_{p}\gg &\frac{2C_{g}eV_{\rm{rms}}^{0}n_{0}}{C_{g}+C_{B}},
		\end{align}
		are satisfied, the counter-rotating terms in Eq.~(\ref{R37}) can be neglected under the rotating-wave approximation. The effective Hamiltonian 
		of the system in the interaction frame becomes
		\begin{align}
			H_{n}=&-K a_{n}^{\dag 2}a_{n}^{2}+\Omega_{p}(a_{n}^{\dag 2}+a_{n}^{2})\cr&+\left[Ja_{n}a_{0}^{\dag}\exp{(i\Delta t)}+{\rm{h.c.}}\right],
		\end{align}
		where $K=2E_{C}/N_{0}^{2}$, $\Omega_{p}=\delta E_{J}\omega_{c}/8E_{J}$,  $J=-i{2C_{g}eV_{\rm{rms}}^{0}n_{0}}/\left({C_{g}+C_{B}}\right)$, and $\Delta=\omega_{0}-\omega_{c}$.
		We have assumed above that the direct coupling between two adjacent KPOs 
		can be neglected because of the long distance between them.
		The total Hamiltonian for the device in Fig.~\ref{figS3}(a) is 
		\begin{align}
			H=&\sum_{n=1}^{N}H_{n}=\sum_{n=1}^{N}-K a_{n}^{\dag 2}a_{n}^{2}+\Omega_{p}(a_{n}^{\dag 2}+a_{n}^{2})\cr &+\left[Ja_{n}a_{0}^{\dag}\exp{(i\Delta t)}+{\rm{h.c.}}\right],
		\end{align}
		which is the Hamiltonian used for our protocol. 
		
		\section{Changing the detuning $\Delta$}\label{App4}
		
		The change of the detuning $\Delta$ can be generally realized using two approaches by: 
			(a) changing the frequency $\omega_{c}$ of the KPOs and (b) inducing a Stark shift for the cavity mode $a_{0}$. 
			Both approaches can be realized by changing the external magnetic flux for transmon qubits.
			A frequency-tunable cavity $a_{0}$ is also a solution for this goal, but it is relatively difficult to experimentally
			change the inductance $L_{r}$ or the capacitance $C_{r}$.

		For the (a) approach, according to Eq.~(\ref{R37}), one can chance the frequency $\omega_{c}=\sqrt{8E_{C}E_{J}/N_{0}}$ for each KPO
		by changing the flux-dependent Josephson energy $E_{J}\rightarrow E'_{J}$.
		Note that, when $E_{J}$ is changed, one needs to adjust the modification $\delta E_{J}\rightarrow \delta E'_{J}$ to satisfy
		$\delta E'_{J}/E'_{J}=\delta E_{J}/E_{J}$,
		so that the two-photon driving amplitude $\Omega_{p}$ remains unchanged.

		%Because the frequency of a transmon qubit is tunable by changing the external magnetic flux, 
		%one can change the generated Stark shift, or, equivalently, change the frequency of the cavity mode $a_{0}$.

		For the (b) approach, we can choose one of the KPOs to be an auxiliary transmon qubit 
		by reducing the number $N_{0}$ of Cooper pairs, e.g., we can assume $N_{0}=1$ for the auxiliary transmon qubit. 
		This auxiliary transmon qubit and the cavity mode $a_{0}$ 
		is designed to be far off-resonant, i.e., their detuning $\Delta_{a}$ is much larger than
		their coupling strength $J_{a}$.
		Then, we arrive at the dispersive
		Hamiltonian
		\begin{align}\label{R41}
			H_{0,a}=\Delta_{s} |e\rangle_{a}\langle e|a_{0}^{\dag}a_{0},
		\end{align}
		where $\Delta_{s}=J_{a}^{2}/\Delta_{a}$ is the Stark shift and $|e\rangle_{a}$ is the excited state of the auxiliary transmon qubit. In this case, when we restrict the auxiliary transmon qubit to be in its ground state,
		Eq.~(\ref{R41}) corresponds to a modification for the frequency of the cavity mode $a_{0}$. 
		The total Hamiltonian becomes
		\begin{align}
			H=&\Delta_{s} a_{0}^{\dag}a_{0}+\sum_{n=1}^{N}-K a_{n}^{\dag 2}a_{n}^{2}+\Omega_{p}\left(a_{n}^{\dag 2}+a_{n}^{2}\right)
			\cr &+\left[Ja_{n}a_{0}^{\dag}\exp{(i\Delta t)}+{\rm{h.c.}}\right].
		\end{align}
		Note that $\Delta_{a}>\Delta$ is tunable by changing the external magnetic flux according to Eq.~(\ref{R37}).
		For $t<\tau$, we assume $\Delta_{a}$ is so large that $\Delta_{s}\rightarrow 0$.
		At time $t=\tau_{m}$, we decrease the detuning $\Delta_{a}$ by changing the external magnetic flux for the auxiliary transmon qubit.
		Then, the detuning between each KPO mode $a_{n}$ and the cavity mode $a_{0}$ becomes $\Delta'=\Delta+\Delta_{s}$.
		This approach has been widely used in quantum measurements, e.g., for the readout of final states.

\begin{acknowledgements}
Y.-H.C. was supported by the Japan Society for the Promotion of Science (JSPS) KAKENHI Grant No.~JP19F19028.
W.Q. was supported in part by the Incentive Research Project of RIKEN.
A.M. was supported by the Polish National Science Centre (NCN)
under the Maestro Grant No.~DEC-2019/34/A/ST2/00081.
F.N. was supported in part by:
Nippon Telegraph and Telephone Corporation (NTT) Research,
the Japan Science and Technology Agency (JST) [via
the Quantum Leap Flagship Program (Q-LEAP),
and the Moonshot R\&D Grant No.~JPMJMS2061],
the Japan Society for the Promotion of Science (JSPS)
[via the Grants-in-Aid for Scientific Research (KAKENHI) Grant No. JP20H00134],
the Army Research Office (ARO) (Grant No.~W911NF-18-1-0358),
the Asian Office of Aerospace Research and Development (AOARD) (via Grant No.~FA2386-20-1-4069), and
the Foundational Questions Institute Fund (FQXi) via Grant No.~FQXi-IAF19-06.
\end{acknowledgements}

%\end{CJK*}
\bibliography{references}

\end{document}